\documentstyle[prb,aps,eqsecnum,epsf,multicol]{revtex}
\newcommand{\bleq}{\ifpreprintsty
                   \else
                   \end{multicols}\vspace*{-3.5ex}{\tiny
                   \noindent\begin{tabular}[t]{c|}
                   \parbox{0.493\hsize}{~} \\ \hline \end{tabular}}
                   \fi}
\newcommand{\eleq}{\ifpreprintsty
                  \else
                   {\tiny\hspace*{\fill}\begin{tabular}[t]{|c}\hline
                    \parbox{0.49\hsize}{~} \\
                    \end{tabular}}\vspace*{-2.5ex}\begin{multicols}{2}
                    \fi}
\newcommand{\bcols}{\ifpreprintsty\else\begin{multicols}{2}\fi}
\newcommand{\ecols}{\ifpreprintsty\else\end{multicols}\fi}

\begin{document} 
  
\draft

\title{Collective modes in a system with two spin-density waves:
the `Ribault' phase of quasi-one-dimensional organic conductors } 

\author{N. Dupuis$^{(1*,2)}$ and Victor M. Yakovenko$^{(2)}$ }
\address{
(1) Laboratoire de Physique des Solides, Universit\'e Paris-Sud, 91405
Orsay, France \\
(2) Department of Physics, University of Maryland, College Park, MD
20742-4111, USA } 
\date{January 21, 2000} 
\maketitle 

\begin{abstract}
We study the long-wavelength collective modes in the magnetic-field-induced
spin-density-wave (FISDW) phases experimentally observed in organic
conductors of the Bechgaard salts family, focusing on phases that exhibit a
sign reversal of the quantum Hall effect (Ribault anomaly). We have recently
proposed that two SDW's coexist in the Ribault phase, as a result of umklapp
processes. When the latter are strong enough, the two SDW's become
circularly polarized (helicoidal SDW's). In this paper, we study the
collective modes which result from the presence of two SDW's. We find two
Goldstone modes, an out-of-phase sliding mode and an in-phase spin-wave
mode, and two gapped modes. The sliding Goldstone mode carries only a
fraction of the total optical spectral weight, which is determined by the
ratio of the amplitude of the two SDW's. In the helicoidal phase, all the
spectral weight is pushed up above the SDW gap. We also point out
similarities with phase modes in two-band, bilayer, or $d+id'$ 
superconductors. We expect our conclusions to hold for generic two-SDW systems.
\end{abstract} 

\pacs{PACS Numbers: 72.15Nj, 73.40Hm, 75.30Fv}   

\bcols

\section{Introduction}

In electron systems with broken symmetries, such as superconductors or 
DW systems, quasi-particle excitations are often gapped, and the only
low-lying excitations are collective modes. The latter thus play a crucial
role in various low-energy properties. 

In an incommensurate SDW system, there are two (gapless) Goldstone modes: a
sliding mode and a spin-wave mode, which result from the spontaneous breaking
of translation symmetry in real space and rotation symmetry in spin space,
respectively.\cite{Gruner,Sengupta99}  Contrary to the case of
superconductors, collective modes in 
DW systems directly couple to external probes and therefore show up in
various experiments. For instance, the sliding mode, which is pinned by
impurities in real systems, can be depinned by a strong electric field. This
leads to non-linear conduction, observed in DW systems. \cite{Gruner} 

The aim of this article is to study long-wavelength collective modes in a
quasi-1D system where 
the low-temperature phase exhibits two SDW's. The presence of two SDW's
gives rise to a rich structure of collective modes, which in principle can
be observed in experiments. \cite{Dupuis99} 

Our results are based on a particular case: the magnetic-field-induced
spin-density-wave (FISDW) phases of the organic conductors of the Bechgaard
salt family. \cite{rev,Heritier84,Poilblanc87,Yakovenko91}
These FISDW phases share common features with standard SDW
phases, but also exhibit remarkable properties like the quantization of the
Hall effect. We have shown that umklapp processes may lead in these systems
to the coexistence of two SDW's with comparable amplitudes, which provides a
possible explanation of the sign reversal of the quantum Hall effect (QHE)
(the so-called Ribault anomaly\cite{Ribault85,Piveteau86}) observed in these
conductors. There are two motivations for studying this particular case. i) The
Bechgaard salts, as a possible candidate for a two-SDW system, present their
own interest. ii) A conductor with two SDW's is in  general not easy to
analyze, even at the mean-field level.\cite{note7} The analysis simplifies
when a strong magnetic field quantizes the electron motion. 

Nevertheless, we expect our conclusions to be quite general and to apply (at
least qualitatively) to other systems with two SDW's. This hope is strongly
supported by the similarities \cite{Dupuis99} that exist between collective
modes in two-SDW conductors and phase modes in two-band,\cite{Legett66}
bilayer\cite{Hwang98} or $d+id'$ (Ref.~\onlinecite{Balatsky99})
superconductors, and, to a lesser extent, plasmon modes
in semiconductor double-well structures. \cite{Dassarma81} These
similarities suggest that collective modes in two-component systems present
generic features that do not depend on the particular case considered.

\subsection{Umklapp processes in quasi-1D SDW systems}

Consider a quasi-1D conductor with a SDW ground state. In presence of
umklapp processes transferring momentum $\bf K$ ($\bf K$ being a vector of
the reciproqual space), spin fluctuations at wave vectors $\bf Q$ and ${\bf
K}-{\bf Q}$ are coupled. Thus, the formation of a SDW at wave vector ${\bf
Q}_1$ will automatically be accompanied by the formation of a second SDW at
wave vector ${\bf Q}_2={\bf K}-{\bf Q}_1$, provided that ${\bf Q}_1\neq {\bf
Q}_2$. The case ${\bf Q}_1={\bf Q}_2={\bf K}/2$ corresponds to a (single)
commensurate SDW. Umklapp processes pin the SDW whose position with respect
to the underlying crystal lattice becomes fixed: the sliding mode is
gapped. For two incommensurate SDW's (${\bf Q}_1\neq {\bf Q}_2$),
\cite{note8} the total spin-density modulation can then be written as
\begin{equation}
\langle {\bf S}({\bf r})\rangle = \sum_{i=1,2} {\bf S}_i \cos({\bf Q}_i\cdot
{\bf r} + \theta_i) ,
\end{equation}
where  ${\bf r}=(na,mb)$ (with $n,m$ integers) denotes the position in real
space ($a$ and $b$ being the lattice spacings along and across the
conducting chains).

Even for ${\bf Q}_1\neq {\bf Q}_2$, the distinction between the two
SDW's may appear somewhat unjustified since 
one cannot distinguish between $\cos({\bf Q}_1\cdot {\bf r})$ and $\cos({\bf
Q}_2\cdot {\bf r})$ when ${\bf r}$ is taken as a discrete variable. However,
umklapp processes do lead to the presence of two non-vanishing order
parameters, $\langle c^\dagger_{{\bf k}\uparrow}c_{{\bf k}+{\bf Q}_1
\downarrow}\rangle$ and  $\langle c^\dagger_{{\bf k}\uparrow} c_{{\bf k}+
{\bf Q}_2\downarrow}\rangle$, in the SDW phase. \cite{Mukhin98} This doubles
the number   
of degrees of freedom of the SDW condensate, which yields for instance twice
as many collective modes (as compared to the case with a single SDW). Thus,
it is natural to speak of two SDW's in the ground state of the
system. Furthermore, we note that $\cos({\bf Q}_1\cdot {\bf r} + \theta_1)$
and  $\cos({\bf Q}_2\cdot {\bf r} + \theta_2)$ are indistinguishable only if
$\theta_1=-\theta_2$ (with, again, $\bf r$ being a discrete variable).
This is precisely the equilibrium condition obtained
by minimizing the mean-field condensation energy (see Sec.~\ref{sec:MF}).
Condensate fluctuations do not in general satisfy the condition
$\theta_1=-\theta_2$, and it is then more appropriate to view these
fluctuations as originating from two different SDW's. 

In a quasi-1D system with a single SDW, the wave vector ${\bf Q}$ of the
spin-density 
modulation is determined by the nesting properties of the Fermi surface:
$E({\bf k}+{\bf Q})\simeq - E({\bf k})$, where $E({\bf k})$ is the energy
with respect to the Fermi level. [For a perfectly nested Fermi surface, one
would have $E({\bf k}+{\bf Q})=- E({\bf k})$.] Umklapp processes are
important only if both ${\bf Q}_1$ and ${\bf Q}_2={\bf K}-{\bf Q}_1$ are
good nesting vectors. Otherwise, one of the two SDW's has a very small
amplitude and can be ignored for any practical purpose. 

For a 2D (or 3D) conductor, the geometry of the Fermi surface appears to be
crucial. Consider the following dispersion law, which is linearized in the
vicinity of the Fermi level:  
\begin{equation}
E_\alpha (k_x,k_y) = v_F(\alpha k_x-k_F)-2t_b\cos(k_yb+\alpha \kappa) +
\cdots 
\label{E0}
\end{equation}
where $k_x$ and $k_y$ are the electron momenta along and across the
conducting chains.
$\alpha=+$ $(-)$ corresponds to right (left) movers with momenta close
to $\alpha k_F$. $v_F$ and $k_F$ are the Fermi velocity and momentum for the
motion along the chains, $t_b$ the interchain transfer integral, and $b$ the
interchain spacing. 
The ellipses in (\ref{E0}) represent small corrections that
generate deviations from perfect nesting. 
$\kappa$ is a parameter which parameterizes the shape of the Fermi
surface. 

Most calculations on quasi-1D SDW systems assume
$\kappa=0$, which corresponds to the Fermi surface shown in
Fig.~\ref{fig:FS}a. There
are two `best' nesting vectors: ${\bf Q}_1\simeq (2k_F,\pi/b-\delta)$ and ${\bf
Q}_2\simeq (2k_F,-\pi/b+\delta)$, where the small correction $\delta$
($|\delta|\ll \pi/b$) is due to deviations from perfect
nesting. \cite{Montambaux85,note9} Since ${\bf Q}_2=(4k_F,0)-{\bf Q}_1$,
these two vectors are coupled by umklapp scattering if the system is
half-filled ($4k_F=2\pi/a$). Two
SDW's with equal amplitudes will form simultaneously at low
temperature. 

Consider now the case $\kappa=\pi/4$ in a half-filled band,
which corresponds to the asymmetric Fermi surface shown in
Fig.~\ref{fig:FS}b. This Fermi surface has been 
proposed as a good approximation to the actual Fermi surface of the
Bechgaard salts. \cite{Ishiguro,note0} The best nesting vector ${\bf Q}_1=
(2k_F,\pi/2b-\delta)$ is now non-degenerate. By umklapp scattering, ${\bf Q}_1$
couples to ${\bf Q}_2=(2k_F,-\pi/2b+\delta)$, which is not a good nesting
vector. At low temperature, two SDW's will form simultaneously, but the one
with wave vector ${\bf Q}_2$ will have a vanishingly small amplitude.

\subsection{The Ribault phase of the Bechgaard salts} 

The organic conductors of the Bechgaard salts family (TMTSF)$_2$X (where
TMTSF stands for tetramethyltetraselenafulvalene) are well known to have
remarkable properties 
in a magnetic field. In three members of this family (X=ClO$_4$, PF$_6$,
ReO$_4$), a moderate magnetic field of a few Tesla destroys the
metallic phase and induces a series of SDW phases
separated by first-order phase transitions. \cite{rev,Heritier84}

According to the so-called quantized nesting model (QNM),
\cite{Heritier84} the formation of the magnetic-field-induced spin-density
waves (FISDW) results from  competition
between the nesting properties of the Fermi surface and the quantization of
the electron motion in a magnetic field. The formation of a  SDW opens a gap,
but leaves closed pockets of electrons and/or holes in the vicinity of the
Fermi surface. In the absence of a magnetic field, these pockets are too large
(due to imperfect nesting) for the SDW phase to be stable. In the 
presence of a magnetic field $H$, they become 
quantized into Landau levels (more precisely Landau subbands). In each FISDW
phase, the SDW wave vector is quantized, ${\bf Q}_N=(2k_F+NG,Q_y)$ with $N$
integer, so that an integer number of Landau subbands are filled. [Here
$G=eHb/\hbar\ll k_F$ and $-e$ is the electron charge.] As a result, the Fermi
level lies 
in a gap between two Landau subbands, the SDW phase is stable, and the Hall
conductivity is quantized: $\sigma _{xy}=-2Ne^2/h$ per one layer of the
TMTSF molecules. \cite{Poilblanc87,Yakovenko91} As the magnetic field
increases, the value of the integer $N$ changes, which leads to a cascade of
FISDW transitions. The QNM predicts the integer $N$ to have always the same
sign. While most of the Hall plateaus are indeed of the same sign, referred
to as positive as convention, a negative QHE is also observed at certain
pressures (the so-called Ribault anomaly). \cite{Ribault85,Piveteau86} 
The most commonly observed negative phases correspond to $N=-2$ and $N=-4$. 

In the Bechgaard salts, a weak dimerization along the chains leads to a
half-filled band. Umklapp processes transferring $4k_F=2\pi/a$ are
allowed. Thus the formation of a SDW at wave vector  ${\bf Q}_N=(2k_F+NG,Q_y)$
will be accompanied by a second SDW at wave vector ${\bf Q}_{\bar N}=(4k_F,0)-
{\bf Q}_N=(2k_F-NG,-Q_y)$, {\it i.e.} there is coexistence of phases $N$ and
$-N$. \cite{Dupuis98} Note that in our notation, ${\bf Q}_{\bar N}$ has both
the signs of $N$ and $Q_y$ reversed compared to ${\bf Q}_N$. As discussed in
the preceding section, actual coexistence may occur only if ${\bf Q}_{\bar N}$
(like ${\bf Q}_N$) is a good nesting vector. This is the case with the Fermi 
surface shown in Fig.~\ref{fig:FS}a (since $Q_y\sim \pi/b$), but not with the 
one shown in Fig.~\ref{fig:FS}b (since  $Q_y\sim \pi/2b$).

In Ref.~\onlinecite{Dupuis98}, we have studied the effect of umklapp 
scattering on the FISDW phases, starting from the Fermi surface shown in 
Fig.~\ref{fig:FS}a. We have shown that for weak umklapp
scattering $Q_y\neq \pi/b$. In that case, the SDW with negative quantum
number $-|N|$ has a vanishing amplitude and can be ignored. However, for $N$
even, there exists a critical value of the umklapp scattering strength above
which the system prefers to form two transversely commensurate SDW's ($Q_y
=\pi/b$). For certain dispersion law,\cite{note10} the SDW with negative
quantum number $-|N|$ has the largest amplitude, which leads to a negative Hall
plateau (Fig.~\ref{fig:Tc}a). Since the umklapp scattering strength is
sensitive to pressure, we 
have suggested that this provides a natural explanation for the negative QHE
(Ribault anomaly) observed in the Bechgaard salts.\cite{Zanchi96} [In the
following, the `negative' phases are referred to as `Ribault' phases.]

It should be noted that this explanation relies on a simple Fermi surface
(Fig.~\ref{fig:FS}a), which does not necessarily provide a good approximation 
to the actual Fermi surface of the Bechgaard salts. With the more realistic 
(according to band calculations) Fermi surface shown
in Fig.~\ref{fig:FS}b, umklapp processes have only a small effect and do not
lead to negative QHE.\cite{Lebed91} Therefore, our explanation of the Ribault
anomaly should be 
taken with caution. For it to  be correct, the parameter $\kappa$ should be
smaller (typically not larger than $\pi/10$) than the value $\pi/4$
predicted by band calculation. In the following, we consider only  the case
$\kappa=0$. 

We have shown in Ref.~\onlinecite{Dupuis98} that 
the SDW's in the Ribault phase are likely to become circularly polarized
(helicoidal SDW's) when the umklapp scattering strength is further
increased (Fig.~\ref{fig:Tc}b). The QHE vanishes in the helicoidal phase. We
will see that the circular polarization also affects the collective modes.

\subsection{Outline of the paper}  

In the next section, we introduce the effective Hamiltonian describing the
FISDW phases. The partition function is written as a functional integral
over a bosonic auxiliary field that describes spin fluctuations. In
Sec.~\ref{sec:MF}, we perform a saddle-point approximation, thus
recovering the mean-field results of Ref.~\onlinecite{Dupuis98}. We obtain the
mean-field propagators and the mean-field particle-hole susceptibilities. In
Sec.~\ref{sec:Seff}, we derive the low-energy effective action of the SDW
phase by taking into account fluctuations of the bosonic auxiliary field
around its saddle-point value. We consider only `phase' fluctuations,
{\it i.e.} sliding and spin-wave collective modes. We do not study amplitude
collective modes, which are gapped and do not couple to phase fluctuations
in the long-wavelength limit.\cite{Gruner}

We find two sliding modes: a (gapless) 
Goldstone mode corresponding to a sliding of the two SDW's in opposite
directions (out-of-phase oscillations), and a gapped mode corresponding to
in-phase oscillations (Sec.~\ref{sec:chmode}). The real part of the
conductivity exhibit two peaks, which reflects the presence of two sliding
modes. The low-energy mode carries only a fraction of the total spectral
weight $\omega_p^2/4$ ($\omega_p$ is the plasma frequency), which is
determined by the ratio of the amplitudes of the two SDW's
(Sec.~\ref{sec:opt}).

The spin-wave modes are studied in
Sec.~\ref{sec:spmode}. There is a Goldstone mode corresponding to in-phase
oscillations of the two SDW's, and a gapped mode corresponding to
out-of-phase oscillations. The spectral function ${\rm Im}\,\chi^{\rm ret}$
is computed in Sec.~\ref{sec:spsp} ($\chi^{\rm ret}$ is the retarded
transverse spin-spin correlation function). Both gapped modes are found to
lie above the mean-field gap in the case of the Bechgaard salts. 

In Sec.~\ref{sec:hel}, we study the
collective modes in the helicoidal phase where the two SDW's are circularly
polarized. For a helicoidal structure, one cannot distinguish between a
uniform spin rotation and a global translation, so that there are only two
phase modes. The Goldstone mode does not contribute to the conductivity and
is therefore a pure spin-wave mode. Thus all the spectral weight in the
conductivity $\sigma(\omega)$ is pushed up above the mean-field gap. Both
modes contribute to the spin-spin correlation function.

It should be noted that the long-wavelength modes are not the only modes of
interest in the FISDW phases. There also exist magneto-rotons at finite wave
vectors ($q_x=G,2G,...$).\cite{Poilblanc87a,Lederer98} These modes are  not
considered in this paper.

In the following, we take $\hbar=k_B=1$.

\section{Model and effective Hamiltonian}
\label{sec:model} 

In the vicinity of the Fermi energy, the electron dispersion law in
the Bechgaard salts is approximated as 
\begin{equation}
E(k_x,k_y)=v_F(|k_x|-k_F)+t_\perp (k_yb) ,
\label{disp}
\end{equation}
where $k_x$ and $k_y$ are the electron momenta along and across the
one-dimensional chains of TMTSF. In Eq.~(\ref{disp}), the longitudinal 
electron dispersion is linearized in $k_x$ in the vicinity of the two
one-dimensional Fermi points $\pm k_F$, and $v_F=2at_a\sin(k_Fa)$ is
the corresponding Fermi velocity ($t_a$ being the transfer integral 
and $a$ the lattice spacing along
the chains). Whenever necessary, we will impose a ultraviolet energy cutoff
$E_0$ to simulate a finite bandwidth. The function $t_\perp(u)$, which
describes the 
propagation in the transverse direction, is periodic: $t_\perp(u)=
t_\perp(u+2\pi)$. It can be expanded in Fourier series as
\begin{eqnarray}
t_\perp(u) &=& -2t_b \cos(u) -2t_{2b} \cos(2u) \nonumber \\ &&
-2t_{3b} \cos(3u) -2t_{4b} \cos(4u) \cdots 
\end{eqnarray}
If we retain only the first harmonic ($t_b$), we obtain a Fermi
surface with perfect nesting at $(2k_F,\pi/b)$. The other harmonics
$t_{2b}, t_{3b} \cdots \ll t_b$ generate deviations from perfect
nesting. They have been introduced in order to keep a realistic
description of the Fermi surface despite the linearization around $\pm
k_F$. In the following, we shall retain only $t_b$, $t_{2b}$ and
$t_{4b}$. $t_{3b}$ does not play an important role and can be
discarded.\cite{Dupuis98} We do not consider the electron dispersion in
the $z$ direction, because it is not important in the following (its
main effect is to introduce a 3D threshold field below which the FISDW
cascade is suppressed\cite{Heritier84}). 

The effect of the magnetic field $H$ along the $z$ direction is taken
into account via the Peierls substitution ${\bf k}\to -i\bbox{\nabla}-e{\bf
A}$. (The charge $e$ is positive since the actual carriers are holes.)
In the gauge ${\bf A}=(0,Hx,0)$ we obtain the non-interacting Hamiltonian
\begin{eqnarray}
{\cal H}_0 &=& \sum_{\alpha,\sigma} \int d^2r\, \hat\psi^\dagger_{\alpha
\sigma} ({\bf r}) [ v_F(-i\alpha \partial_x-k_F) \nonumber \\ &&
+ t_\perp(-ib\partial_y-Gx) +\sigma h]\hat\psi_{\alpha\sigma}({\bf r}) .
\label{ham1}
\end{eqnarray}
Here $\hat\psi_{\alpha\sigma}({\bf r})$ is a fermionic operator for right
($\alpha =+$) and left ($\alpha=-$) moving particles. $\sigma=+$ ($-$)
for up (down) spin. We use the notation ${\bf r}=(x,mb)$ ($m$ integer)
and $\int d^2r=b\sum_m \int dx$. $G=eHb$ is a magnetic wave vector and
$h=\mu_BH$ is the Zeeman energy (we assume the electron gyromagnetic factor to
be equal to two). Diagonalizing the Hamiltonian (\ref{ham1}) we obtain the
eigenstates and eigenenergies 
\begin{eqnarray}
\phi^\alpha_{\bf k}({\bf r}) &=& \frac{1}{\sqrt{L_xL_y}} 
e^{i{\bf k}\cdot {\bf r} + i(\alpha/\omega_c) T_\perp (k_yb-Gx)} , 
\nonumber \\ 
\epsilon_{\alpha\sigma}({\bf k}) & \equiv & \epsilon_{\alpha\sigma}(k_x) 
= v_F(\alpha k_x-k_F) + \sigma h ,
\label{basis}
\end{eqnarray}
where $L_xL_y$ is the area of the system, $\omega_c=v_FG$, and
\begin{equation}
T_\perp (u) = \int_0^u du' \, t_\perp (u') . 
\end{equation}
In the chosen gauge the energy depends only on
$k_x$, {\it i.e.} the dispersion law is one-dimensional. This reflects the
localization of the electron motion in the transverse direction, which is at
the origin of the QNM (see Ref.~\onlinecite{Dupuis98} for a further
discussion). Note that contrary to $k_y$, the quantum number 
$k_x$ is not the momentum since the operator 
$\partial_x$ does not commute with the Hamiltonian. 

The interacting part of the Hamiltonian contains two terms
corresponding to forward ($g_2$) and umklapp ($g_3$) scattering: 
\begin{eqnarray}
\label {ham2}
{\cal H}_{\rm int} &=& \frac{g_2}{2} \sum_{\alpha,\sigma,\sigma'} 
\int d^2r \,
\hat\psi^\dagger_{\alpha \sigma} ({\bf r})
\hat\psi^\dagger_{\bar\alpha \sigma'} ({\bf r}) 
\hat\psi_{\bar\alpha \sigma'} ({\bf r}) 
\hat\psi_{\alpha \sigma} ({\bf r}) \nonumber \\ && + 
\frac{g_3}{2} \sum_{\alpha,\sigma} 
\int d^2r \, e^{-i\alpha 4k_Fx}
\hat\psi^\dagger_{\bar \alpha \sigma} ({\bf r})
\hat\psi^\dagger_{\bar\alpha \bar\sigma} ({\bf r}) 
\hat\psi_{\alpha \bar\sigma} ({\bf r}) 
\hat\psi_{\alpha \sigma} ({\bf r}) , \nonumber \\ && 
\end{eqnarray}
where $\bar\alpha=-\alpha$ and $\bar\sigma=-\sigma$. 
For repulsive interaction $g_2,g_3\geq 0$. We shall assume that umklapp
scattering is `weak': $g_3<g_2$. We do not consider backward
scattering ($g_1$) since it does not play an important role in the
following. The interaction strength is best parametrized by dimensionless
coupling constants $\tilde g_2=g_2N(0)$ and $\tilde g_3=g_3N(0)$, where
$N(0)=1/\pi v_Fb$ is the density of states per spin. 

In a mean-field theory, the cutoff $E_0$ is of the order of the bandwidth
$t_a$. It is however well known that mean-field theory completely neglects
fluctuations and cannot be directly applied to quasi-1D systems where the
physics, at least at high temperature, is expected to be
one-dimensional. The general wisdom is that there is at low temperature a
dimensional crossover from a 1D to a 2D (or 3D) regime. 
\cite{Bourbon91,Kishina99} In Bechgaard salts (at ambient pressure),
this dimensional crossover occurs via
coherent single-particle interchain hopping. Although the experimental
assignment of the crossover temperature is still under debate
(see for instance Refs.~\onlinecite{Moser98,Schartz98,Fertey99}), the 
success of the QNM in explaining the phase diagram of the compounds
(TMTSF)$_2$PF$_6$ and (TMTSF)$_2$ClO$_4$ provides a clear evidence of the
relevance of the Fermi surface at low temperature. In the 3D regime, a
mean-field theory is justified provided that the parameters of the theory are
understood as effective parameters (renormalized by 1D
fluctuations). \cite{Bourbon91} $E_0$
is then smaller than the bare bandwidth ($\sim t_a$) and corresponds to the
renormalized transverse bandwidth. For the same reason, the interaction
amplitudes $g_2$ and $g_3$ can be different than their bare values. 
The Hamiltonian [Eqs.~(\ref{ham1}) and (\ref{ham2})] should therefore be
understood as an effective low-energy Hamiltonian.  

Since collective modes are best studied within a
functional integral formalism, we write the partition function as
$Z=\int {\cal D}\psi^*{\cal D}\psi \,e^{-S_0-S_{\rm int}}$ where
$\psi^{(*)}$ is an anticommuting Grassmann variable. The actions $S_0$
and $S_{\rm int}$ are given by
\begin{eqnarray}
&& S_0 = \int d\tau \, \Bigl \lbrace \sum_{\alpha,\sigma} \int d^2r\, 
\psi^*_{\alpha \sigma} ({\bf r},\tau) \partial_\tau
\psi_{\alpha \sigma} ({\bf r},\tau)
+{\cal H}_0[\psi^*,\psi] \Bigr \rbrace , \nonumber \\  
&& S_{\rm int} = -\frac{1}{2}  \sum_{\alpha,\alpha',\sigma} \int d^2r\,d\tau\, 
O^*_{\alpha\sigma}({\bf r},\tau) \hat g_{\alpha\alpha'}({\bf r})
O_{\alpha'\sigma}({\bf r},\tau) , 
\end{eqnarray}
where the imaginary time $\tau$ varies between 0 and $\beta=1/T$. We
have introduced the composite field
\begin{equation}
O_{\alpha\sigma}({\bf r},\tau) =
\psi^*_{\bar\alpha \bar\sigma} ({\bf r},\tau)
\psi_{\alpha \sigma} ({\bf r},\tau) 
\end{equation}
and the matrix 
\begin{equation}
\hat g({\bf r}) = 
\left (
\begin{array}{lr}
g_2  & g_3e^{i4k_Fx} \\ g_3e^{-i4k_Fx} & g_2 \\
\end{array}
\right ) .
\end{equation}

Introducing a complex auxiliary field $\Delta_{\alpha\sigma}({\bf
r},\tau)$, we decouple $S_{\rm int}$ by means of a Hubbard-Stratonovitch
transformation. This leads to the action
\begin{eqnarray}
S &=& S_0 + \int d^2r\,d\tau\, \Delta^\dagger_\uparrow({\bf r},\tau)\hat 
g({\bf r}) \Delta_\uparrow({\bf r},\tau) \nonumber \\ && - \sum_{\sigma}
\int d^2r\, d\tau\, \Delta^\dagger_\sigma({\bf r},\tau)\hat g({\bf r})
O_\sigma({\bf r},\tau) .
\label{action1} 
\end{eqnarray}
We use the spinor notation $\Delta_\sigma=(\Delta_{+\sigma},
\Delta_{-\sigma})^T$ and  $O_\sigma=(O_{+\sigma},O_{-\sigma})^T$.
Since $\Delta_{\alpha\sigma}$ couples to the field $O_{\alpha\sigma} ({\bf
r},\tau) = O^*_{\bar\alpha\bar\sigma} ({\bf r},\tau)$, it satisfies the
constraint $\Delta_{\alpha\sigma}({\bf r},\tau) =
\Delta^*_{\bar\alpha\bar\sigma} ({\bf r},\tau)$. Note that the action
(\ref{action1}) maintains spin-rotation invariance around the magnetic field 
axis. In the FISDW phase, because of the Zeeman term, the magnetization is
perpendicular to the magnetic field axis, so that  the 
only spin-wave mode corresponds to rotation around the $z$ axis. In zero 
magnetic field, a different approach should be used in order to maintain the
full SU(2) spin symmetry (see Ref.~\onlinecite{Sengupta99}).

\section{Mean-field theory}
\label{sec:MF} 

In this section we look for a mean-field solution corresponding to a
phase with two sinusoidal ({\it i.e.} linearly polarized) SDW's:
\begin{eqnarray}
\langle S_x({\bf r}) \rangle &=& \sum_{p=\pm} M_{pN} \cos(\phi_{pN})
\cos({\bf Q}_{pN}\cdot {\bf r} + \theta_{pN}) \nonumber \\ 
\langle S_y({\bf r}) \rangle &=& \sum_{p=\pm} M_{pN} \sin(\phi_{pN})
\cos({\bf Q}_{pN}\cdot {\bf r} + \theta_{pN}) \nonumber \\ 
\langle S_z({\bf r}) \rangle &=& 0 ,
\label{SDWdef}
\end{eqnarray}
where $S_\nu({\bf r})=\sum_{\alpha\sigma\sigma'}
\psi^*_{\bar\alpha\sigma} ({\bf r}) \tau^\nu_{\sigma\sigma'}
\psi_{\alpha\sigma'} ({\bf r})$ is the spin-density operator and
$\tau^\nu$ ($\nu=x,y,z$) the Pauli matrices. Because of the Zeeman
coupling with the magnetic field, the SDW's are polarized in the
$(x,y)$-plane. The variable $\phi_{pN}$ determines the polarization axis,
while $\theta_{pN}$ gives the position of the SDW's with respect to the
underlying crystal lattice. 

We assume that the external parameters (magnetic field, pressure...) are
such that the system is in the Ribault phase, characterized by a negative
QHE and the coexistence of two SDW's with comparable amplitudes. We choose
the sign of $N$ such that $N$ refers to the SDW with the largest amplitude
($M_N\geq M_{\bar N}$). [$N$ is even and negative in the Ribault phase.] 

The mean-field solution corresponds to a saddle-point
approximation with a static auxiliary field \cite{note3} 
\begin{equation}
\Delta_{\alpha\sigma}({\bf r})= \langle O_{\alpha\sigma}({\bf r},\tau)
\rangle = \sum_p \Delta^{(pN)}_{\alpha\sigma}
e^{i\alpha {\bf Q}_{pN}\cdot {\bf r} }. 
\end{equation}
The relation between $\Delta_{\alpha\sigma}$ and $\langle O_{\alpha\sigma}
\rangle$ results from the stationarity condition of the saddle-point action
(see Eq.~(\ref{action11}) below). 
Because of the constraint $\Delta^*_{\alpha\sigma}({\bf r})=\Delta
_{\bar\alpha\bar\sigma}({\bf r})$, the order parameters satisfy
$\Delta^{(pN)*}_{\alpha\sigma}=\Delta^{(pN)} _{\bar\alpha\bar\sigma}$.
Among the eight complex order parameters, only four are therefore
independent and sufficient to characterize the SDW phase. 
The order parameters $\Delta^{(pN)}_{\alpha\sigma}=  |
\Delta^{(pN)}_{\alpha\sigma}| e^{ i\varphi^{(pN)}_{\alpha\sigma}}$ are
related to the spin-density modulation (\ref{SDWdef}) by
\begin{eqnarray}
|\Delta^{(pN)}_{\alpha\sigma}| &=& \frac{M_{pN}}{4}, \nonumber \\ 
\theta_{pN} &=& \frac{1}{2}\Bigl ( \varphi^{(pN)}_{+\uparrow} -
\varphi^{(pN)}_{-\uparrow} \Bigr ), \nonumber \\ 
\phi_{pN} &=& -\frac{1}{2}\Bigl ( \varphi^{(pN)}_{+\uparrow} +
\varphi^{(pN)}_{-\uparrow} \Bigr ).
\label{SDWdef1}
\end{eqnarray}
Since by convention $M_N\geq M_{\bar N}$, we have
$|\Delta^{(N)}_{\alpha\sigma}| \geq |\Delta^{(\bar N)}_{\alpha\sigma}|$. 
Note that the condition $\Delta^{(pN)*}_{\alpha\sigma}= \Delta^{(pN)}
_{\bar\alpha\bar\sigma}$ imposes $\varphi^{(pN)}_{\bar\alpha\bar\sigma}=
-\varphi^{(pN)}_{\alpha\sigma}$. 

The mean-field action is given by
\bleq
\begin{eqnarray}
S_{\rm MF} &=& \beta \int d^2r\, \Delta^\dagger_\uparrow ({\bf r})
\hat g({\bf r}) \Delta _\uparrow ({\bf r}) -
\sum_\alpha \int d^2r\,d\tau \int d^2r'\,d\tau'\, 
(\psi^*_{\alpha\uparrow}({\bf r},\tau) , \,
\psi^*_{\bar\alpha\downarrow} ({\bf r},\tau) ) \nonumber \\ && \times 
\left (
\begin{array}{lr}
G_{\alpha\uparrow}^{(0)-1}({\bf r},\tau;{\bf r}',\tau')  & 
\delta({\bf r}-{\bf r}') \delta(\tau-\tau')
\tilde\Delta_{\alpha\uparrow}({\bf r}) \\ 
\delta({\bf r}-{\bf r}') \delta(\tau-\tau')
\tilde\Delta^*_{\alpha\uparrow}({\bf r})
& G_{\bar\alpha\downarrow}^{(0)-1}({\bf r},\tau;{\bf r}',\tau')  \\
\end{array}
\right )
\left (
\begin{array}{l}
\psi_{\alpha\uparrow}({\bf r}',\tau') \\ 
\psi_{\bar\alpha\downarrow}({\bf r}',\tau')
\end{array} \right ) .
\label{action11}
\end{eqnarray}
Here $G_{\alpha\uparrow}^{(0)}({\bf r},\tau;{\bf r}',\tau')$ is the
Green's function in the absence of interaction ($g_2=g_3=0$). 
The effective potential $\tilde\Delta_{\alpha\sigma}({\bf r})$
acting on the electrons is defined by
\begin{equation}
\tilde\Delta_{\alpha\sigma}({\bf r})=g_2 \Delta_{\alpha\sigma}({\bf
r}) + g_3 e^{i\alpha4k_Fx} \Delta_{\bar\alpha\sigma}({\bf r}),
\end{equation}
{\it i.e.} $\tilde\Delta_{\sigma}({\bf r})=\hat g({\bf
r})\Delta_{\sigma}({\bf r})$. Rewriting the action in the basis of the
eigenstates $\phi^\alpha_{\bf k}$ of ${\cal H}_0$ [Eq.~(\ref{basis})], we
obtain 
\begin{eqnarray}
S_{\rm MF} &=& \beta \int d^2r\, \Delta^\dagger_\uparrow ({\bf r})
\hat g({\bf r}) \Delta _\uparrow ({\bf r}) \nonumber \\ && -
\sum_{\alpha,{\bf k},{\bf k}',\omega}
(\psi^*_{\alpha\uparrow}({\bf k},\omega) , \,
\psi^*_{\bar\alpha\downarrow} ({\bf k},\omega) ) 
\left (
\begin{array}{lr}
\delta_{{\bf k},{\bf k}'}(i\omega -\epsilon_{\alpha\uparrow}(k_x))  & 
\tilde\Delta_{\alpha\uparrow}({\bf k},{\bf k}') \\ 
\tilde\Delta_{\bar\alpha\downarrow}({\bf k},{\bf k}')
& \delta_{{\bf k},{\bf k}'}(i\omega -\epsilon_{\bar\alpha\downarrow}(k_x))  \\
\end{array}
\right )
\left (
\begin{array}{l}
\psi_{\alpha\uparrow}({\bf k}',\omega) \\ 
\psi_{\bar\alpha\downarrow}({\bf k}',\omega)
\end{array} \right ) .
\label{action12}
\end{eqnarray}
\eleq 
We denote by
$\omega=2\pi T(n+1/2)$ ($n$ integer) fermionic Matsubara frequencies.
We emphasize that here ${\bf k}$ and ${\bf k}'$ refer to the quantum numbers
of the eigenstates $\phi^\alpha_{\bf k}$ of ${\cal H}_0$.
We have introduced the (particle-hole) pairing amplitudes 
\begin{eqnarray}
\tilde \Delta_{\alpha\sigma}({\bf k},{\bf k}') &=&  \int d^2r\,
\phi^{\alpha *}_{\bf k}({\bf r}) \phi^{\bar\alpha}_{{\bf k}'}({\bf r})
\tilde \Delta_{\alpha\sigma}({\bf r}) \nonumber \\ &=&
\delta_{k_y',k_y-\pi/b} \sum_p(g_2 \Delta^{(pN)}_{\alpha\sigma} + g_3
\Delta^{(\bar pN)}_{\bar\alpha\sigma}) \nonumber \\ && \times
\sum_{n=-\infty}^{\infty} I_n e^{i\alpha n(k_yb-\pi/2)}
\delta_{k_x',k_x-\alpha Q_x^{(pN)}+\alpha nG} , \nonumber \\
\label{pa1}
\end{eqnarray}
where $I_n\equiv I_n(q_y=\pi/b)$. The coefficients $I_n(q_y)$, defined by
\begin{equation}
I_n(q_y) = \int_0^{2\pi}  \frac{du}{2\pi}\, 
e^{inu+(i/\omega_c)[T_\perp(u+q_yb/2)+T_\perp(u-q_yb/2)]} ,
\label{IN}
\end{equation}
are well know in the QNM. They depend on the transverse dispersion
law $t_\perp(k_yb)$ and measure the degree of perfect nesting of
the Fermi surface. For a perfect nesting at wave vector $(2k_F,\pi/b)$, one
would have $I_n(\pi/b)=\delta_{n,0}$, while in the generic situation
(imperfect nesting) $|I_n(q_y)|<1$. \cite{note5}
The convention $M_{\bar N}\leq M_N$ implies $|I_{\bar N}|
\leq |I_N|$. Note that $k_x$ is coupled not only to $k_x-\alpha Q_x^{(pN)}$
but also to $k_x-\alpha Q_x^{(pN)}+\alpha nG$ [Eq.~(\ref{pa1})]. This
reflects the fact that $k_x$ is not the momentum in the gauge ${\bf
A}=(0,Hx,0)$. 

From now on, we shall use the quantum limit approximation (QLA) (also
known as the single gap approximation\cite{Heritier84}). This approximation,
which is valid when $\omega_c$ is much larger than the temperature and the SDW
gap, consists in retaining only the most singular (particle-hole) pairing
channels. In the metallic phase, these channels present the
logarithmic singularity $\sim\ln(E_0/T)$ (which is characteristic of a SDW
system with perfect nesting). This singularity results from pairing between
electron and hole states of the same energy and is responsible for the
opening of a gap at the Fermi level. While the QLA strongly underestimates
the transition temperature and the value of the gap at low temperature, and
also neglects gaps opening above and below the Fermi level, it is an
excellent approximation when calculating the collective modes (we shall
come back to this point). \cite{Poilblanc87a} 

Thus, in the QLA, $\tilde\Delta_{\alpha\sigma}({\bf
k},{\bf k}')$ is not zero only if $\epsilon_{\alpha\sigma}(k_x)=
-\epsilon_{\bar\alpha\bar\sigma}(k_x')$, which requires
$k_x'=k_x-\alpha 2k_F$. Only the term with $n=pN$ survives in
Eq.~(\ref{pa1}), which yields 
\begin{eqnarray}
\tilde\Delta_{\alpha\sigma}({\bf k},{\bf k}') &=& \delta_{{\bf k}',{\bf
k}-\alpha {\bf Q}_0} \tilde\Delta_{\alpha\sigma}(k_y) , \nonumber \\ 
\tilde\Delta_{\alpha\sigma}(k_y) &=& \sum_p \tilde\Delta^{(pN)}_{\alpha\sigma}
e^{i\alpha pN(k_yb-\pi/2)} ,
\label{DTky}
\end{eqnarray}
where ${\bf Q}_0=(2k_F,\pi/b)$.
We have introduced the new order parameters 
$\tilde\Delta^{(pN)}_{\alpha\sigma}$ related to 
$\Delta^{(pN)}_{\alpha\sigma}$ by 
\begin{eqnarray}
\tilde\Delta^{(pN)}_{\alpha\sigma}  &=& I_{pN} (g_2
\Delta^{(pN)}_{\alpha\sigma} + g_3
\Delta^{(\bar pN)}_{\bar\alpha\sigma}) =  \tilde\Delta^{(pN)*}
_{\bar\alpha\bar\sigma}  , \nonumber \\ 
\Delta^{(pN)}_{\alpha\sigma}  &=&
\frac{g_2 I_{\bar pN} \tilde\Delta^{(pN)}_{\alpha\sigma} - g_3 I_{pN} 
\tilde\Delta^{(\bar pN)}_{\bar\alpha\sigma}}{(g_2^2-g_3^2)I_NI_{\bar N}}. 
\label{deltaT}
\end{eqnarray}
In the QLA, the mean-field action therefore reduces to a $2\times 2$ matrix 
(as in BCS theory):
\bleq 
\begin{eqnarray}
S_{\rm MF} &=& \beta \int d^2r\, \Delta^\dagger_\uparrow ({\bf r})
\hat g({\bf r}) \Delta _\uparrow ({\bf r}) \nonumber \\ && -
\sum_{\alpha,{\bf k},\omega}
(\psi^*_{\alpha\uparrow}({\bf k},\omega) , \,
\psi^*_{\bar\alpha\downarrow} ({\bf k}-\alpha{\bf Q}_0,\omega) ) 
\left (
\begin{array}{lr}
i\omega -\epsilon_{\alpha\uparrow}(k_x)  & 
\tilde\Delta_{\alpha\uparrow}(k_y) \\ 
\tilde\Delta^*_{\alpha\uparrow}(k_y)
& i\omega +\epsilon_{\alpha\uparrow}(k_x)  \\
\end{array}
\right )
\left (
\begin{array}{l}
\psi_{\alpha\uparrow}({\bf k},\omega) \\ 
\psi_{\bar\alpha\downarrow}({\bf k}-\alpha{\bf Q}_0,\omega)
\end{array} \right ) , 
\label{Sqla}
\end{eqnarray}
\eleq
using $\epsilon_{\bar\alpha\downarrow}(k_x-\alpha 2k_F)=-
\epsilon_{\alpha\uparrow}(k_x)$. It is remarkable that within the QLA the
mean-field action can still be written as a $2\times 2$ matrix. The presence
of a second SDW changes the expression of the pairing amplitude (which
becomes $k_y$-dependent), but not the fact that the state $({\bf k},-)$
couples only to $({\bf k}+{\bf Q}_0,+)$. This is not true in zero magnetic
field where the presence of a second SDW leads to complicated mean-field
equations. \cite{note7}

\subsection{Ground-state energy}

The mean-field action being Gaussian, we can integrate out the fermion
fields to obtain the ground-state condensation energy (per unit area), 
$\Delta E=-(T\ln Z)/L_xL_y-E_N$, where $T\to 0$ and $E_N=-N(0)E_0^2$ is the
normal state energy: 
\begin{eqnarray}
\Delta E &=& \sum_\alpha \Biggl \lbrace \sum_p
\frac{\Delta^{(pN)*}_{\alpha\uparrow} \tilde
\Delta^{(pN)}_{\alpha\uparrow}}{I_{pN}} + \frac{N(0)}{2}
|\tilde\Delta^{(\bar N)}_{\alpha\uparrow}|^2 \nonumber \\ && 
- \frac{N(0)}{2} \sum_p |\tilde\Delta^{(pN)}_{\alpha\uparrow}|^2 
\Biggl ( \frac{1}{2} + \ln
\frac{2E_0}{|\tilde\Delta^{(N)}_{\alpha\uparrow}|} \Biggr ) \Biggr \rbrace .
\label{DE}
\end{eqnarray}
To obtain (\ref{DE}), we have explicitly taken into account the fact that
$|\tilde\Delta^{(N)}_{\alpha\sigma}| \geq |\tilde\Delta^{(\bar N)}
_{\alpha\sigma}|$ and $|\Delta^{(N)}_{\alpha\sigma}| \geq
|\Delta^{(\bar N)}_{\alpha\sigma}|$. The latter inequality follows from
$M_N\geq M_{\bar N}$, while the former is derived below Eq.~(\ref{ratio1}). 
Using the gap equation $\partial\Delta E/\partial \tilde\Delta
^{(pN)*}_{\alpha\uparrow}=0$, we can rewrite the condensation energy as
\begin{equation}
\Delta E = -\frac{N(0)}{4} \sum_{\alpha,p} 
|\tilde\Delta^{(pN)}_{\alpha\uparrow}|^2 ,
\end{equation}
a form which is reminiscent of the BCS-like expression of a SDW system with
perfect nesting ($\Delta E=-N(0)|\Delta|^2/2$ with $\Delta$ the SDW gap). As
stated above, the QLA amounts to taking into account only the `perfect
nesting component' of the particle-hole pairing. 

The requirement that $\Delta E$ is minimal with respect to variation
of the order parameters leads to constraints on the phases of the
complex order parameters $\tilde\Delta_{\alpha\sigma}^{(pN)}= 
|\tilde\Delta_{\alpha\sigma}^{(pN)}| e^{i \tilde 
\varphi^{(pN)}_{\alpha\sigma}}$: 
\begin{equation}
\tilde \varphi^{(N)}_{\alpha\uparrow}-\tilde \varphi^{(\bar N)}_{\bar
\alpha\uparrow} = 
\left \lbrace 
\begin{array}{lr}
0   & {\rm if} \,\,\,  I_NI_{\bar N} > 0, \\
\pi & {\rm if} \,\,\,  I_NI_{\bar N} < 0. 
\end{array} \right.
\label{phiT1}
\end{equation}
For $t_{4b}=0$, $I_{\bar N}=(-1)^{N/2}I_N$. When $t_{4b}$ is finite but small,
${\rm sgn}(I_NI_{\bar N})=(-1)^{N/2}$. 
Using Eqs.~(\ref{deltaT}) and (\ref{phiT1}), we then obtain
\begin{equation}
\frac{\Delta^{(\bar N)}_{\bar \alpha\uparrow}}{\Delta^{(N)}
_{\alpha\uparrow}}  = \frac{|\tilde\Delta^{(\bar N)}_{\bar\alpha\uparrow}/
\tilde\Delta^{(N)}_{\alpha\uparrow}
|-r|I_{\bar N}/I_N|}{|I_{\bar N}/I_N|-r|\tilde\Delta^{(\bar
N)}_{\bar\alpha\uparrow}/ \tilde\Delta^{(N)}_{\alpha\uparrow}|}, 
\label{ratio1}
\end{equation}
Eq.~(\ref{ratio1}), together with $|\Delta^{(\bar
N)}_{\bar\alpha\sigma} /\Delta^{(N)}_{\alpha\sigma}|=M_{\bar N}/M_N\leq 1$ 
and $|I_{\bar N}/I_N|\leq 1$,
implies $|\tilde\Delta^{(\bar
N)}_{\bar\alpha\sigma} /\tilde\Delta^{(N)}_{\alpha\sigma}|\leq 1$. 
It also shows that $\Delta^{(\bar N)}_{\bar \alpha\uparrow}/ \Delta^{(N)}
_{\alpha\uparrow}$ is a real number, so that the phases
$\varphi^{(N)}_{\alpha\uparrow}$ and $\varphi^{(\bar N)}_{\bar
\alpha\uparrow}$ are identical modulo $\pi$:
\begin{equation}
\varphi^{(N)}_{\alpha\uparrow}=\varphi^{(\bar N)}_{\bar \alpha\uparrow}
\,\,\, [\pi].
\label{phiT2}
\end{equation}
From (\ref{SDWdef1}) and (\ref{phiT2}), we then deduce 
\begin{equation}
\phi_N = \phi_{\bar N} , \,\,\,\,\,\,
\theta_N = - \theta_{\bar N} . 
\end{equation}
In the mean-field ground state, the two SDW's have the same
polarization axis ($\phi_N=\phi_{\bar N}$). The condition $\theta_N
=-\theta_{\bar N}$ means that the two SDW's can be displaced in
opposite directions without changing the energy of the system. They can 
therefore be centered either on the lattice sites or on the bonds.  For 
$\theta_N=\theta_{\bar N}=0$, both SDW's are centered on the sites, while for
$\theta_N=-\theta_{\bar N}=\pi/2$, they are centered on the bonds. The 
condition $\theta_N+\theta_{\bar N}=0$ is related to the pinning that would 
occur for a commensurate SDW. Indeed, for a single SDW with wave vector 
$(2k_F,\pi/b)$, it becomes $2\theta=0$ where $\theta$ is the phase of the SDW. 
\cite{Gruner} 

This degeneracy of the ground state results from rotational invariance
around the $z$-axis in spin space,\cite{note1} and translational invariance in
real space. The latter  holds in the FISDW states, since the SDW's are
incommensurate with respect to the crystal lattice.\cite{note2}
According to the mean-field analysis, and in agreement with general
symmetry considerations, we therefore expect two (gapless) Goldstone
modes: a spin-wave mode corresponding to a uniform rotation around the
$z$-axis of the common polarization axis, and a sliding mode
corresponding to a displacement of the two SDW's in opposite
directions. 

Without loss of generality, we can choose the phases $\tilde
\varphi^{(pN)}_{\alpha\sigma}$ and $\varphi^{(pN)}_{\alpha\sigma}$
equal to 0 or $\pi$. The order parameters $\tilde\Delta
^{(pN)}_{\alpha\sigma}$ and $\Delta ^{(pN)}_{\alpha\sigma}$ are then
real (positive or negative) numbers. Furthermore, since the
constraints obtained by minimizing the free energy relate  $\tilde
\varphi^{(N)}_{\alpha\sigma}$ ($\varphi^{(N)}_{\alpha\sigma}$)
only to  $\tilde \varphi^{(\bar N)}_{\bar\alpha\sigma}$
($\varphi^{(\bar N)}_{\bar\alpha\sigma}$), we can also impose $\tilde
\varphi^{(N)}_{\alpha\sigma}=\tilde \varphi^{(
N)}_{\bar\alpha\sigma}$ and  $\varphi^{(N)}_{\alpha\sigma}=\varphi^{(
N)}_{\bar\alpha\sigma}$. With such a choice, 
\begin{equation}
\tilde\Delta^{(pN)}_{\alpha\sigma}\equiv \tilde\Delta_{pN}, \,\,\,\, 
\Delta^{(pN)}_{\alpha\sigma}\equiv \Delta_{pN}
\end{equation}
are real (positive or negative) numbers independent of
$\alpha$ and $\sigma$. $\theta_{pN}=0$ and $\phi_N=\phi_{\bar N}=0$ or
$\pi$. The SDW's are polarized along the $x$ axis. Introducing the notations 
\begin{eqnarray}
r = \frac{g_3}{g_2} , & \,\,\,  & \zeta =\frac{I_{\bar N}}{I_N}, \nonumber \\
\gamma = \frac{\Delta_{\bar N}}{\Delta_N} ,
& \,\,\,  & \tilde\gamma = \frac{\tilde\Delta_{\bar N}}{\tilde\Delta_N} ,
\end{eqnarray}
(with $|\gamma|=M_{\bar N}/M_N$), we deduce from (\ref{deltaT})
\begin{equation}
\tilde \gamma = \zeta \frac{\gamma +r}{1+r\gamma} , \,\,\, 
\gamma = \frac{\tilde \gamma -r \zeta}{\zeta -r\tilde \gamma}. 
\label{gam2t}
\end{equation}
Note that $\tilde\gamma$ and $\zeta$ have the same sign [see (\ref{phiT1})], 
and $|\gamma|,|\tilde\gamma|,r,|\zeta|\leq 1$. In the Ribault phase,
$|\gamma| \simeq |\tilde\gamma|$. \cite{Dupuis98} Both SDW's have the same
amplitude when $|\zeta|=1$. In that case $\tilde\gamma=\zeta$ and $\gamma=1$.

\subsection{Mean-field propagators}
\label{sec:MFpro} 

The normal and anomalous Green's functions are defined by
\begin{eqnarray}
G_{\alpha\sigma}({\bf r},{\bf r}',\omega) &=& 
-\langle \psi_{\alpha\sigma}({\bf r},\omega) 
\psi^*_{\alpha\sigma}({\bf r}',\omega) \rangle \nonumber \\ &=& 
\sum_{{\bf k},{\bf k}'}  G_{\alpha\sigma}({\bf k},{\bf k}',\omega)
\phi^\alpha_{\bf k}({\bf r}) \phi^{\alpha *}_{{\bf k}'}({\bf r}') , 
\nonumber \\
F_{\alpha\sigma}({\bf r},{\bf r}',\omega) &=& 
-\langle \psi_{\alpha\sigma}({\bf r},\omega) 
\psi^*_{\bar\alpha\bar\sigma}({\bf r}',\omega) \rangle \nonumber \\ &=& 
\sum_{{\bf k},{\bf k}'}  F_{\alpha\sigma}({\bf k},{\bf k}',\omega)
\phi^\alpha_{\bf k}({\bf r}) \phi^{\bar\alpha *}_{{\bf k}'}({\bf r}') .
\label{propa1}
\end{eqnarray}
From (\ref{Sqla}), we deduce that
\begin{eqnarray}
G_{\alpha\sigma}({\bf k},{\bf k}',\omega) &=& \delta_{{\bf k},{\bf
k}'} G_{\alpha\sigma}({\bf k},\omega) , \nonumber \\ 
F_{\alpha\sigma}({\bf k},{\bf k}',\omega) &=& \delta_{{\bf k}',{\bf
k}-\alpha {\bf Q}_0} F_{\alpha\sigma}({\bf k},\omega) ,
\end{eqnarray}
with 
\begin{eqnarray}
G_{\alpha\sigma}({\bf k},\omega) &=& 
\frac{-i\omega-\epsilon_{\alpha\sigma}(k_x)}{\omega^2+\epsilon^2
_{\alpha\sigma}(k_x) +|\tilde\Delta_{\alpha\sigma}(k_y)|^2} ,
\nonumber \\ 
F_{\alpha\sigma}({\bf k},\omega) &=& 
\frac{\tilde\Delta_{\alpha\sigma}(k_y)}{\omega^2+\epsilon^2
_{\alpha\sigma}(k_x) +|\tilde\Delta_{\alpha\sigma}(k_y)|^2} .
\end{eqnarray}
The excitation energies are given by $\pm E_{\alpha\sigma}({\bf k})$, where
\begin{equation}
E_{\alpha\sigma}({\bf k}) = \sqrt{\epsilon_{\alpha\sigma}^2(k_x) +
|\tilde\Delta_{\alpha\sigma}(k_y)|^2} . 
\label{energy1}
\end{equation} 
Writing
$\tilde\Delta_{\alpha\uparrow}^{(pN)} = |\tilde\Delta_{\alpha\uparrow}
^{(pN)}| e^{i\tilde \varphi_{\alpha\sigma}^{(pN)}}$,
we obtain (for $N$ even) 
\begin{eqnarray}
|\tilde\Delta_{\alpha\sigma}(k_y)|^2 &=& \sum_p
|\tilde\Delta_{pN}|^2 + 2
|\tilde\Delta_N\tilde\Delta_{\bar N}| 
\nonumber \\ && \times 
\cos(2\alpha Nk_yb + \tilde\varphi^{(N)}_{\alpha\sigma} - 
\tilde\varphi^{(\bar N)}_{\alpha\sigma} ) .
\end{eqnarray}
The quasi-particle excitation gap $2|\tilde\Delta_{\alpha\sigma}(k_y)|$
depends on the transverse momentum $k_y$ (Fig.~\ref{fig:QP}). 
When both SDW's have the same amplitude, $|\tilde\Delta_N|=
|\tilde\Delta_{\bar N}|$, the spectrum becomes gapless since $\tilde\Delta
_{\alpha\sigma}(k_y)$ vanishes at some points on the Fermi surface. 

We have shown in Ref.~\onlinecite{Dupuis98} that the occurrence of helicoidal
phases prevents the spectrum from becoming gapless. Indeed, the stability of
the sinusoidal phase requires
\begin{equation}
\frac{{\rm min}_{k_y} |\tilde\Delta_{\alpha\sigma}(k_y)|}{{\rm max}_{k_y} 
|\tilde\Delta_{\alpha\sigma}(k_y)|} = 
\frac{1-|\tilde\gamma|}{1+|\tilde\gamma|} 
\gtrsim 0.32 ,
\label{crit}
\end{equation} 
{\it i.e.} $|\tilde\gamma|\lesssim 0.52$. 
Eq.~(\ref{crit}) is deduced from a Ginzburg-Landau expansion of the free
energy (valid only close to the transition line) and may be slightly changed
at low temperature. 

In the helicoidal phase, there is no dispersion of the gap with respect to
$k_y$ (see Sec.~\ref{sec:hel}). Since the spectrum becomes gapless for
$|\tilde\gamma|=1$, it is 
natural that the system, above a certain value of $|\tilde\gamma|$, prefers
to form helicoidal SDW's in order to lower the free energy by opening a
large gap on the whole Fermi surface.

\subsection{Mean-field susceptibilities}
\label{sec:chi}

The results reported in this section are derived in detail in Appendix
\ref{app:chi}. We study the mean-field susceptibilities (which will be
useful in Sec.~\ref{sec:Seff}) 
\begin{eqnarray}
\bar \chi_{\alpha\sigma,\alpha'\sigma'}({\bf r},\tau;{\bf r}',\tau') 
&=&\langle O_{\alpha\sigma}({\bf r},\tau)O^*_{\alpha'\sigma'}({\bf r}',\tau') 
\rangle \nonumber \\ && - \langle O_{\alpha\sigma}({\bf r},\tau)\rangle \langle
O^*_{\alpha'\sigma'}({\bf r}',\tau') \rangle ,
\end{eqnarray} 
where the mean value $\langle \cdots \rangle$ is taken with the mean-field
action $S_{\rm MF}$ [Eq.~(\ref{Sqla})]. Since the latter is Gaussian, we
immediately deduce the only two non-vanishing components
\begin{eqnarray}
\bar\chi_{\alpha\sigma,\alpha\sigma}({\bf r},\tau;{\bf r}',\tau') &=& -
G_{\alpha\sigma}({\bf r},\tau;{\bf r}',\tau') 
G_{\bar\alpha\bar\sigma}({\bf r}',\tau';{\bf r},\tau) , \nonumber \\ 
\bar\chi_{\alpha\sigma,\bar\alpha\bar\sigma}({\bf r},\tau;{\bf r}',\tau')
&=& -  
F_{\alpha\sigma}({\bf r},\tau;{\bf r}',\tau') 
F_{\alpha\sigma}({\bf r}',\tau';{\bf r},\tau) . \nonumber \\ 
\end{eqnarray} 

At $T=0$ and for $q_y=0$, the Fourier transformed susceptibilities are given 
by (Appendix \ref{app:chi})
\bleq
\begin{eqnarray}
\bar\chi_{+\uparrow,+\uparrow}({\bf q}+ {\bf Q}_{pN},{\bf q}
+ {\bf Q}_{pN},\omega_\nu) &=& \frac{N(0)}{2} I_{pN}^2 
\Biggl [ \ln \frac{2E_0}{|\tilde\Delta_N|} -\frac{1}{2} 
- \frac{\omega_\nu^2+v_F^2q_x^2}{6(1-\tilde\gamma^2)\tilde\Delta_N^2}  
\Biggr ] , \nonumber \\ 
\bar\chi_{+\uparrow,+\uparrow}({\bf q}+ {\bf Q}_{pN},{\bf q}
+ {\bf Q}_{\bar pN},\omega_\nu) &=& -\frac{N(0)}{4} I_NI_{\bar N} 
\tilde\gamma \Biggl [ 1 -  \frac{\omega_\nu^2+v_F^2q_x^2}{3(1-\tilde\gamma^2)
\tilde\Delta_N^2}  \Biggr ] , \nonumber \\ 
\bar\chi_{+\uparrow,-\downarrow}({\bf q}+ {\bf Q}_{pN},{\bf q}
- {\bf Q}_{pN},\omega_\nu) &=& -\delta_{p,+}
\frac{N(0)}{4} I_N^2 (1-\tilde\gamma^2)
\Biggl [ 1 - \frac{\omega_\nu^2+v_F^2q_x^2}{6(1-\tilde\gamma^2)
\tilde\Delta_N^2} \Biggr ] ,
\nonumber \\ 
\bar\chi_{+\uparrow,-\downarrow}({\bf q}+ {\bf Q}_{pN},{\bf q}
- {\bf Q}_{\bar pN},\omega_\nu) &=& -\frac{N(0)}{4} I_NI_{\bar N} 
\tilde\gamma  ,
\label{chi1}
\end{eqnarray}
\eleq
where  $\omega_\nu=2\pi T\nu$ ($\nu$ integer) is a bosonic Matsubara
frequency.
Eqs.~(\ref{chi1}) are valid in the low-energy limit $v_F^2q_x^2,\omega_\nu^2
\ll \tilde\Delta^2_N(1-\tilde\gamma^2)$. We have used the QLA (see Appendix
\ref{app:chi}), whose 
validity is discussed in detail in Ref.\onlinecite{Poilblanc87a}. Although
the QLA does not predict accurately the value of the mean-field gaps $\tilde
\Delta_{pN}$ and $\Delta_{pN}$, it is an excellent approximation of the
low-energy properties (such as the long-wavelength collective modes) once the
values of $\tilde \Delta_{pN}$ and $\Delta_{pN}$ are known. Since the
absolute value of the latter do not play an important role for our purpose,
the QLA turns out to be a perfect mean to compute the collective modes.

\section{Low-energy effective action}
\label{sec:Seff}

In this section we derive the low-energy effective action determining the
sliding and spin-wave modes. We do not  consider amplitude modes which are
gapped and do not couple to `phase' fluctuations in the long-wavelength
limit. We consider only longitudinal ({\it i.e.} with the wave vector
parallel to the chains)  
fluctuations ($q_y=0$). These fluctuations are of particular interest since
they couple to an electric field applied along the chains (which is a common
experimental situation): see Sec.~\ref{sec:opt} on the optical
conductivity. Including a finite $q_y$ would allow to obtain the effective
mass of the collective modes in the transverse direction. Such a calculation
is however much more involved and will not be attempted here. 

The low-energy effective action is derived by studying fluctuations of the
auxiliary field 
$\Delta_{\alpha\sigma}({\bf r},\tau)$ around its saddle-point value
$\Delta_{\alpha\sigma}({\bf r})$. We write $\Delta_{\alpha\sigma}({\bf
r},\tau) = \Delta_{\alpha\sigma}({\bf r}) + \eta_{\alpha\sigma}({\bf r},\tau)$
and calculate the effective action to quadratic order in the fluctuating
field $\eta$. The fermionic action (\ref{action1}) can be rewritten as
\begin{eqnarray}
S[\psi^*,\psi] &=& S_{\rm MF} [\psi^*,\psi] 
- \sum_\sigma \int d^2r\, d\tau\, \eta^\dagger_\sigma ({\bf
r},\tau) \tilde O_\sigma({\bf r},\tau) 
\nonumber \\ && + \int d^2r\, d\tau\, [\eta^\dagger_\uparrow ({\bf
r},\tau)\hat g({\bf r})\eta_\uparrow ({\bf r},\tau) 
 \nonumber \\ && + (\eta^\dagger_\uparrow ({\bf r},\tau)
\tilde\Delta_\uparrow({\bf r}) + {\rm c.c.}) ] .
\end{eqnarray}
Integrating out the fermions, we obtain to quadratic order in $\eta$ the
effective action
\begin{eqnarray}
S[\eta^*,\eta] &=& \frac{1}{2} \int d^2r\,d\tau\, \int d^2r'\,d\tau'\, 
\sum_{\alpha,\alpha',\sigma,\sigma'} \eta^*_{\alpha\sigma} ({\bf r},\tau) 
\nonumber \\ && \times 
\Bigl [ \delta_{\sigma,\sigma'} \delta ({\bf r}-{\bf r}') \delta
(\tau-\tau') \hat g_{\alpha\alpha'}({\bf r}) \nonumber \\ && -
\tilde\chi_{\alpha\sigma,\alpha'\sigma'}({\bf r},\tau;{\bf r}',\tau') 
\Bigr ] \eta_{\alpha'\sigma'}({\bf r}',\tau') ,
\label{action2}
\end{eqnarray}
where the susceptibility $\tilde\chi$ is defined by (\ref{chiT}). There is
no linear term since we expand around the saddle point. Considering
phase fluctuations only, 
\begin{equation}
\Delta_{\alpha\sigma}({\bf r},\tau) = \sum_p \Delta_{pN} 
e^{i\alpha {\bf Q}_{pN}\cdot {\bf r} + i \varphi^{(pN)}_{\alpha\sigma} ({\bf
r},\tau) } ,
\end{equation}
we have 
\begin{equation}
\eta_{\alpha\sigma}({\bf r},\tau) = i \sum_p \Delta_{pN}
e^{i\alpha {\bf Q}_{pN}\cdot {\bf r}} \varphi^{(pN)}_{\alpha\sigma} ({\bf
r},\tau)
\label{etaT} 
\end{equation}
to lowest order in the (real) phase field $\varphi^{(pN)}_{\alpha\sigma} ({\bf
r},\tau)$. Note that the property $\Delta^*_{\alpha\sigma}({\bf r},\tau) = 
\Delta_{\bar\alpha\bar\sigma}({\bf r},\tau)$ imposes $\varphi^{(pN)}
_{\bar\alpha\bar\sigma} ({\bf r},\tau)=-\varphi^{(pN)}_{\alpha\sigma} ({\bf
r},\tau)$. Using this latter condition, we deduce from (\ref{action2}) the
effective action for the phase 
\begin{eqnarray}
S[\varphi] &=& \sum_{p,p'} \Delta_{pN} \Delta_{p'N}
\sum_{\alpha,\alpha',\tilde q} 
\varphi^{(pN)}_{\alpha\uparrow}(-\tilde q) 
\varphi^{(p'N)}_{\alpha'\uparrow}(\tilde q)  
\nonumber \\ && \times
\Bigl [ g_2 \delta_{\alpha,\alpha'}\delta_{p,p'} +
g_3 \delta_{\alpha,\bar\alpha'}\delta_{p,\bar p'} + A^{pp'}_{\alpha\alpha'}
(\tilde q) \Bigr ] ,
\end{eqnarray}
where $\tilde q=(q_x,\omega_\nu)$, and 
\begin{eqnarray}
A^{pp'}_{\alpha\alpha'}(\tilde q) &=&
-\tilde\chi_{\alpha\uparrow,\alpha'\uparrow}({\bf q}+\alpha {\bf Q}_{pN}, 
{\bf q}+\alpha'{\bf Q}_{p'N},\omega_\nu) \nonumber \\ && + 
\tilde\chi_{\alpha\uparrow,\bar\alpha'\downarrow}({\bf  q}+\alpha {\bf Q}
_{pN}, {\bf q}-\alpha'{\bf Q}_{p'N},\omega_\nu) . 
\label{Sphi1}
\end{eqnarray}
In Eq.~(\ref{Sphi1}) ${\bf q}=(q_x,q_y=0)$. 
Using Eq.~(\ref{chiT1}), and introducing charge and spin variables as in the 
mean-field theory [see 
Eq.~(\ref{SDWdef1})],
\begin{eqnarray}
\theta_{pN}({\bf r},\tau) &=& \frac{1}{2} \Bigl
(\varphi^{(pN)}_{+\uparrow}({\bf r},\tau) - \varphi^{(pN)}_{-\uparrow}({\bf
r},\tau)\Bigr ) , \nonumber \\ 
\phi_{pN}({\bf r},\tau) &=&  -\frac{1}{2} \Bigl
(\varphi^{(pN)}_{+\uparrow}({\bf r},\tau) + \varphi^{(pN)}_{-\uparrow}({\bf
r},\tau)\Bigr ), 
\end{eqnarray}
we obtain a decoupling of the sliding and spin-wave modes, which are
determined by the actions
\begin{equation}
S[\theta] = \frac{1}{2} \sum_{\tilde q} (\theta_N(-\tilde q), \,
\theta_{\bar N}(-\tilde q)) {\cal D}_{\rm ch}^{-1}(\tilde q) 
\left (
\begin{array}{l}
\theta_N(\tilde q) \\ 
\theta_{\bar N}(\tilde q)
\end{array} \right ) ,
\label{Stheta} 
\end{equation}
\begin{equation}
S[\phi] = \frac{1}{2} \sum_{\tilde q} (\phi_N(-\tilde q), \,
\phi_{\bar N}(-\tilde q)) {\cal D}_{\rm sp}^{-1}(\tilde q) 
\left (
\begin{array}{l}
\phi_N(\tilde q) \\ 
\phi_{\bar N}(\tilde q)
\end{array} \right ) ,
\label{Sphi}
\end{equation}
with
\bleq
\begin{eqnarray}
{\cal D}_{\rm ch}^{-1}(\tilde q) &=& 4g_2 \Delta_N^2 
\left (
\begin{array}{lr}
1-c_{++}-r^2c_{--}+2rc_{+-}  &  
-\gamma r (1-c_{++}-c_{--})-\gamma(1+r^2)c_{+-}  \label{Dprop1} 
\\ 
-\gamma r (1-c_{++}-c_{--})-\gamma(1+r^2)c_{+-}  &
\gamma^2(1-c_{--}-r^2c_{++}) + 2\gamma^2 r c_{+-} 
\end{array} \right ) , \\ 
{\cal D}_{\rm sp}^{-1}(\tilde q) &=& 4g_2 \Delta_N^2 
\left (
\begin{array}{lr}
1-c_{++}-r^2c_{--}-2rc_{+-}  &  
\gamma r (1-c_{++}-c_{--})-\gamma(1+r^2)c_{+-}  
\\ 
\gamma r (1-c_{++}-c_{--})-\gamma(1+r^2)c_{+-}  &
\gamma^2(1-c_{--}-r^2c_{++}) - 2\gamma^2 r c_{+-} 
\end{array} \right ) .
\label{Dprop2}
\end{eqnarray}
\eleq
The coefficients $c_{pp'}\equiv c_{pp'}(\tilde q)$ are linear combinations of
the mean-field susceptibilities $\bar\chi$ and are defined in Appendix
\ref{app:chi}. 

The effective actions [Eqs.~(\ref{Stheta}) and (\ref{Sphi})] have been
obtained by 
expanding the auxiliary field with respect to the phase fluctuations
$\varphi_{\alpha\sigma}^{(pN)}({\bf r},\tau)$ [Eq.~(\ref{etaT})]. While this
procedure turns out to be correct at zero temperature, an alternative and
more rigorous approach consists in performing a chiral rotation of the
fermion fields, which allows to directly expand with respect to
$\partial_\tau\varphi_{\alpha\sigma}^{(pN)}({\bf r},\tau)$ and 
$\bbox{\nabla}\varphi_{\alpha\sigma}^{(pN)}({\bf
r},\tau)$. \cite{Sengupta99,Yak98}
The drawback of this approach is that one has to calculate the non-trivial
Jacobian of the chiral transformation (the so-called chiral anomaly).

\section{Sliding modes}
\label{sec:chmode}

The dispersion of the sliding modes is obtained from ${\rm det} {\cal D}
_{\rm ch}^{-1}(\tilde q)=0$. From (\ref{Dprop1}), we deduce
\begin{eqnarray}
{\rm det}{\cal D}_{\rm ch}^{-1}(\tilde q) &=& 16g_2^2 \Delta_N^4 \gamma^2
(1-r^2) [(1-c_{++})(1-c_{--}) \nonumber \\ && -r^2c_{++}c_{--} 
+2rc_{+-}-(1-r^2)c_{+-}^2 ] . \nonumber \\ && 
\label{ch1}
\end{eqnarray}
The gap equation (\ref{gap3}) shows that ${\rm det} {\cal D}_{\rm ch}^{-1}
(\tilde q=0)=0$, which indicates that there is a gapless mode with vanishing
energy in the limit of long-wavelength ($q_x\to 0$). 
Eq.~(\ref{ch1}) simplifies into
\begin{eqnarray}
\delta c_{++} \frac{r\zeta}{\tilde\gamma} +\delta c_{--} \frac{r\tilde
\gamma}{\zeta} -2r \delta c_{+-} && \nonumber \\ 
- (1-r^2)(\delta c_{++}\delta c_{--}-\delta c^2_{+-}) &=& 0. 
\label{ch2}
\end{eqnarray}

\subsection{Goldstone mode}

Eq.~(\ref{ch2}) admits the solution $\omega_\nu^2+v_F^2q_x^2=0$. After
analytical continuation to real frequencies ($i\omega_\nu \to \omega +i0^+$),
we obtain a mode with a linear dispersion law (Fig.~\ref{fig:cm}) 
\begin{eqnarray}
\omega &=& v_Fq_x , \\ 
\theta_N(q_x,\omega) &=& - \theta_{\bar N}(q_x,\omega) .
\end{eqnarray}
As expected from the mean-field analysis (Sec.~\ref{sec:MF}), the
Goldstone mode corresponds to an out-of-phase oscillation of the two SDW's.
We do not find any renormalization of the mode velocity because we have 
not taken into account long-wavelength charge 
fluctuations. The latter couple to the sliding modes and renormalize the 
velocity to $v_F(1+g_4N(0))^{1/2}$, where $g_4$ is the forward scattering
strength. \cite{Gruner} 
In order to obtain this velocity renormalization, it would have been necessary 
to include forward scattering in the interaction Hamiltonian 
and to introduce an auxiliary field for the long-wavelength charge 
fluctuations.\cite{Sengupta99}

\subsection{Gapped mode}

The second mode obtained from (\ref{ch2}) corresponds to a gapped mode with
dispersion law (Fig.~\ref{fig:cm})
\begin{eqnarray}
\omega^2 &=& v_F^2q_x^2 + \omega_0^2 ,  \\ 
\omega_0^2 &=& \frac{12}{\tilde g_2} \frac{r}{1-r^2} 
\frac{3+5\tilde\gamma^2}{3\tilde\gamma^2} \Biggl | 
\frac{\tilde\Delta_N\tilde\Delta_{\bar N}}{I_NI_{\bar N}} \Biggr | .
\label{om0}
\end{eqnarray}
The oscillations of this mode satisfy 
\begin{equation}
\frac{\theta_N}{\theta_{\bar N}} = 
\frac{r\zeta-\tilde\gamma}{r\tilde \gamma -\zeta}  
\frac{r(3+\tilde\gamma^2+4\tilde\gamma\zeta)^2+(1-r)^24\zeta\tilde\gamma(3+
\tilde\gamma^2) }{(3+\tilde\gamma^2+4r\tilde\gamma\zeta)^2} . 
\end{equation} 
If we consider the case where the two SDW's have the same amplitude (which
is not physical because of the appearance of the helicoidal phase), which
occurs when $\tilde\gamma=\zeta=\pm 1$, we obtain the simple result
$\theta_N =\theta_{\bar N}$.

Using the physical parameters of the Bechgaard salts, we find that
$\omega_0$ is generally larger than the order parameters $|\tilde\Delta_{\pm
N}|$, so that the gapped sliding mode appears above the quasi-particle
excitation gap (in general within the first Landau subband above the Fermi
level). We therefore expect this mode to be strongly damped due to the
coupling with the quasi-particle excitations. Note that Eq.~(\ref{om0})
giving $\omega_0$ is correct only if  $\omega_0\ll |\tilde\Delta_N|
(1-\tilde\gamma^2)^{1/2}$,
since it was obtained using expressions of the susceptibilities $c_{pp'}$
valid in the limit $\omega_\nu^2,v_F^2q_x^2\ll \tilde\Delta^2_N
(1-\tilde\gamma^2)$. 

It is worth pointing out that the two sliding modes bear some similarities
with phase modes occurring in two-band\cite{Legett66} or bilayer\cite{Hwang98}
superconductors. [This also holds for the spin-wave
modes (Sec.~\ref{sec:spmode}) and the collective modes of the helicoidal
phase (Sec.~\ref{sec:hel}).] $g_3/g_2$ plays the same role as the ratio
between the interband (or interlayer) and intraband (or intralayer) coupling
constants. There are also analogies with phase modes in $d+id'$ 
superconductors, and, to some extent, with plasmon modes
occurring in conducting bilayer systems.\cite{Dassarma81} While the
corresponding phase modes in superconducting systems have not yet been
observed, plasmon modes in semiconductor double-well structure have been
observed recently via inelastic-light-scattering
experiments. \cite{Kainth98}

\section{Optical conductivity} 
\label{sec:opt}

At $T=0$ there are no thermally excited carriers above the SDW gap. The
response to a low-frequency electric field is then entirely due to
fluctuations of the 
SDW condensate. Since the electric field does not couple to the
amplitude modes, we only have to consider the phase fluctuations studied in the
preceding sections. In the next section, we determine the charge ($\rho_{\rm
DW}$) and current ($j_{\rm DW}$) fluctuations induced by phase
fluctuations of the condensate. In Sec.~\ref{sec:current}, we calculate
the current-current correlation function $\langle j_{\rm DW}j_{\rm DW}
\rangle$ which determines the conductivity.  

\subsection{Charge and current operators} 

In this section we calculate explicitly the charge induced by phase
fluctuations of the condensate. The current is then obtained from the
continuity equation.

We consider a source field $A_0({\bf r},\tau)$ which couples to the
(long-wavelength) charge density fluctuations. This adds to the fermionic
action the contribution
\begin{equation}
\Delta S[\psi^*,\psi] = \int d^2r\, d\tau\, A_0({\bf r},\tau) \sum_{\alpha,
\sigma} \psi^*_{\alpha\sigma}({\bf r},\tau) \psi_{\alpha\sigma}({\bf r},\tau).
\end{equation}
The charge operator is then obtained from the action by functional
differentiation with respect to the external field $A_0$:
\begin{equation}
\rho ({\bf r},\tau) = \frac{\delta S}{\delta A_0({\bf r},\tau)} \Biggr
|_{A_0=0} .
\label{op1}
\end{equation}
Following Sec.~\ref{sec:model}, we introduce the auxiliary field $\Delta
_{\alpha\sigma}({\bf r},\tau)$ and integrate out the fermions. This leads to
the action 
\begin{eqnarray}
S[\Delta^*,\Delta,A_0] &=& \int d^2r\, d\tau\, \Delta^\dagger_\uparrow 
({\bf r},\tau) \hat g({\bf r}) \Delta_\uparrow({\bf r},\tau) \nonumber \\ && 
- \sum_\alpha {\rm Tr} \ln (-{\cal G}^{-1}_\alpha +\hat A_0 ) ,
\end{eqnarray}
where 
\bleq
\begin{equation}
{\cal G}^{-1}_\alpha ({\bf r},\tau;{\bf r}',\tau') =
\left (
\begin{array}{lr} 
G^{(0)-1}_{\alpha\uparrow} ({\bf r},\tau;{\bf r}',\tau') & 
\delta ({\bf r}-{\bf r}') \delta (\tau-\tau') 
\tilde\Delta_{\alpha\uparrow}({\bf r},\tau) \\
\delta ({\bf r}-{\bf r}') \delta (\tau-\tau') 
\tilde\Delta^*_{\alpha\uparrow}({\bf r},\tau) & 
G^{(0)-1}_{\bar\alpha\downarrow} ({\bf r},\tau;{\bf r}',\tau') 
\end{array} 
\right ) ,  
\end{equation}
\eleq 
\begin{equation}
\hat A_0 ({\bf r},\tau;{\bf r}',\tau') = \delta ({\bf r}-{\bf r}') 
\delta (\tau-\tau') A_0({\bf r},\tau) \hat 1. 
\end{equation}
We denote by $\hat 1$ the $2\times 2$ unit matrix. 
Expanding the action with respect to $A_0$, we obtain
\begin{equation}
S[\Delta^*,\Delta,A_0] = S[\Delta^*,\Delta] + \sum_\alpha {\rm Tr} 
[{\cal G}_\alpha \hat A_0] +O(A_0^2) ,
\end{equation}
where $S[\Delta^*,\Delta]$ is the action without the source field. Here
${\rm Tr}$ denotes the trace with respect to time, space and matrix indices. 
From (\ref{op1}), we then obtain the following expression of the charge
operator
\begin{equation}
\rho ({\bf r},\tau) = \sum_\alpha {\rm tr}\, {\cal G}_\alpha ({\bf r},\tau;
{\bf r},\tau) ,
\end{equation}
where $\rm tr$ denotes the trace with respect to the matrix indices
only ({\it i.e.} ${\rm Tr}\,\hat O=\int d^2r\,d\tau\, {\rm tr}\,\hat O({\bf
r},\tau; {\bf r},\tau)$ for any operator $\hat O$). 

To calculate $\rho$, we write 
\begin{eqnarray}
\tilde\Delta_\sigma({\bf r},\tau) &=& \tilde\Delta_\sigma({\bf r}) 
+ \tilde\eta_\sigma ({\bf r},\tau) , \nonumber \\ 
\tilde\eta_\sigma ({\bf r},\tau) &=& \hat g({\bf r}) \eta_\sigma({\bf r},\tau).
\label{etaT1}
\end{eqnarray} 
Then we have
\begin{equation}
{\cal G}^{-1}_\alpha = {\cal G}^{{\rm MF}-1}_\alpha - \Sigma_\alpha, 
\end{equation}
\begin{eqnarray}
\Sigma_\alpha ({\bf r},\tau;{\bf r}',\tau') &=& -
\delta ({\bf r}-{\bf r}') \delta (\tau-\tau') \nonumber \\ && \times 
\left ( 
\begin{array}{lr} 
0 & \tilde\eta_{\alpha\uparrow} ({\bf r},\tau) \\ 
\tilde\eta^*_{\alpha\uparrow} ({\bf r},\tau) & 0 
\end{array}
\right ) ,
\end{eqnarray} 
where ${\cal G}^{\rm MF}$ is the mean-field propagator.  
The charge operator is given by 
\begin{eqnarray} 
\rho ({\bf r},\tau) &=& \sum_\alpha {\rm tr}\, {\cal G}_\alpha^{\rm MF} ({\bf
r},\tau; {\bf r},\tau) \nonumber \\ && 
+ \sum_\alpha {\rm tr} [{\cal G}_\alpha^{\rm MF} 
\Sigma_\alpha {\cal G}_\alpha^{\rm MF}]({\bf r},\tau;{\bf r},\tau) +
O(\tilde\eta^2) .
\label{op2}
\end{eqnarray}
The first term in the rhs of (\ref{op2}) is the uniform charge density
$\rho_0$ in the mean-field state. The other terms correspond to charge
fluctuations $\rho_{\rm DW}$ induced by the condensate fluctuations. To
lowest order in phase fluctuations, we obtain
\begin{equation}
\rho_{\rm DW} ({\bf r},\tau) = \sum_\alpha  {\rm tr} [{\cal G}_\alpha^{\rm
MF} \Sigma_\alpha {\cal G}_\alpha^{\rm MF}]({\bf r},\tau;{\bf r},\tau) .
\label{op3}
\end{equation}

Considering only phase fluctuations [Eq.~(\ref{etaT})], we obtain 
(Appendix \ref{app:rho})  
\begin{equation}
\rho_{\rm DW} ({\bf r},\tau) = \frac{1}{\pi b} \partial_x \tilde\theta_N
({\bf r},\tau) ,
\label{charge1}
\end{equation}
where 
\begin{eqnarray}
\tilde\theta_N ({\bf r},\tau) &=& \frac{1}{2} \Bigl ( 
\tilde\varphi^{(N)}_{+\uparrow} ({\bf r},\tau) - 
\tilde\varphi^{(N)}_{-\uparrow} ({\bf r},\tau) \Bigr ) , \\
\tilde\varphi^{(pN)}_{\alpha\sigma} ({\bf r},\tau) &=& \frac{g_2 \Delta _{pN} 
\varphi^{(pN)}_{\alpha\sigma} ({\bf r},\tau) +g_3 \Delta_{\bar pN} 
\varphi^{(\bar pN)}_{\bar\alpha\sigma} ({\bf r},\tau)}{g_2 \Delta _{pN} +
g_3 \Delta_{\bar pN}} . 
\label{phiT3}
\end{eqnarray}
Note that $\rho_{\rm DW} ({\bf r},\tau)$ depends only on $\tilde\theta_N({\bf
r},\tau)$, {\it i.e.} on the largest component of the effective potential
$\tilde\Delta_{\alpha\sigma}({\bf r},\tau)$. 

Using the continuity equation (in imaginary time) 
\begin{equation}
ie\partial_\tau \rho_{\rm DW} ({\bf r},\tau) 
+ \partial_x j_{\rm DW} ({\bf r},\tau) =0 ,
\end{equation}
we obtain the current fluctuations induced by the phase fluctuations:
\begin{equation}
j_{\rm DW} ({\bf r},\tau) = -\frac{ie}{\pi b} \partial_\tau
\tilde\theta_N ({\bf r},\tau) .
\label{cur1}
\end{equation}
Note that the current $j_{\rm DW}$ is parallel to the chains. Condensate
fluctuations do not induce current fluctuations in the transverse direction. 

This simple result [Eqs.~(\ref{charge1}) and (\ref{cur1})] can be understood as
follows. For a system with a single SDW, the charge fluctuation $\rho_{\rm
DW} =\partial_x\theta/\pi b$ is obtained by requiring the SDW gap to remain
tied to the
Fermi level. Let us briefly recall this argument. A fluctuating SDW
potential $\Delta_{\alpha\sigma}(x,\tau)=\Delta_{\alpha\sigma} e^{i\alpha
2k_Fx +i\varphi _{\alpha\sigma}(x,\tau)}\simeq \Delta _{\alpha\sigma}
e^{i\alpha 2k_Fx +i x \partial_x 
\varphi _{\alpha\sigma}(x,\tau)}$ couples the state $(k_x,\downarrow)$ to 
$(k_x+2k_F+\partial_x\varphi_{+\uparrow}(x,\tau),\uparrow)$, and 
$(k_x,\uparrow)$ to $(k_x+2k_F-\partial_x\varphi_{-\uparrow}(x,\tau),
\downarrow)$. [For simplicity, we neglect the transverse direction.] 
The resulting gap will open at the Fermi level only if the
system modifies its density in such a way that 
\begin{eqnarray}
\delta k_{F+\uparrow}(x,\tau) + \delta k_{F-\downarrow}(x,\tau) &=& \partial_x
\varphi_{+\uparrow}(x,\tau), \nonumber \\  
\delta k_{F+\downarrow}(x,\tau) + \delta k_{F-\uparrow}(x,\tau) &=& -
\partial_x \varphi_{-\uparrow}(x,\tau) .
\label{kF} 
\end{eqnarray}
Here we denote by $\alpha (k_F+\delta k_{F\alpha\sigma}(x,\tau))$ the Fermi
wave vector on the $(\alpha\sigma)$ branch of the spectrum. This Fermi wave
vector is time and space dependent because of the density
fluctuations. Eqs.~(\ref{kF}) imply the charge density variation
\begin{eqnarray}
\delta \rho (x,\tau) &=& \frac{1}{2\pi b} \sum _\sigma \Bigl (\delta
k_{F+\sigma} (x,\tau) + \delta k_{F-\sigma} (x,\tau)\Bigr ) 
\nonumber \\ &=& \frac{1}{2\pi b} \partial_x \Bigl ( \varphi_{+\uparrow}
(x,\tau) - \varphi_{-\uparrow}(x,\tau) \Bigr ) \nonumber \\ 
&=& \frac{1}{\pi b} \partial_x
\theta (x,\tau). 
\end{eqnarray}
Let us  now go back to the case of interest. We should consider the
effective potential $\tilde\Delta_{\alpha\sigma}({\bf r},\tau)$, since it is 
$\tilde\Delta_{\alpha\sigma}({\bf r},\tau)$ (and not
$\Delta_{\alpha\sigma}({\bf r},\tau)$) which couples to the electrons. In
presence of two SDW's, it is not possible to satisfy (\ref{kF}) for the two
components ($N$ and $-N$) of the effective potential. Nevertheless, it is
natural to assume that the largest gap remains tied to the Fermi level. 
This immediately yields Eq.~(\ref{charge1}), and, from the continuity equation,
Eq.~(\ref{cur1}). [The latter can also be obtained by a similar argument
without using the continuity equation.\cite{Gruner}] In conclusion, 
Eqs.~(\ref{charge1}) and (\ref{cur1}) express the fact that the largest gap
remains tied to the Fermi level.  

Finally, we can express the charge and current fluctuations directly in
terms of $\theta_N$ and $\theta_{\bar N}$:
\begin{eqnarray}
\rho_{\rm DW}({\bf r},\tau) &=& \frac{1}{\pi b} 
\frac{\partial_x\theta_N({\bf r},\tau)-r\gamma\partial_x\theta_{\bar N}({\bf
r},\tau)}{1+r\gamma} , 
\nonumber \\ 
j_{\rm DW}({\bf r},\tau) &=& - \frac{ie}{\pi b} 
\frac{\partial_\tau\theta_N({\bf r},\tau)-r\gamma\partial_\tau\theta_{\bar
N}({\bf r},\tau)}{1+r\gamma} . 
\end{eqnarray}

\subsection{Current-current correlation function}
\label{sec:current}

In this section we calculate the correlation function 
\begin{equation}
\Pi (\omega_\nu) = \langle j_{\rm DW}(\omega_\nu)j_{\rm DW}(-\omega_\nu)
\rangle ,
\end{equation}
where $j_{\rm DW}(\omega_\nu)\equiv j_{\rm DW}({\bf q}=0,\omega_\nu)$ is the
Fourier transform of the current operator given by (\ref{cur1}). We have
\begin{eqnarray}
\Pi (\omega_\nu) &=& \frac{e^2\omega_\nu^2}{\pi^2b^2(1+\gamma r)^2}
\nonumber \\ && \times [{\cal D}_{\rm ch}^{++}(\omega_\nu) 
+r^2\gamma^2 {\cal D}_{\rm ch}^{--}(\omega_\nu)-2r\gamma 
{\cal D}_{\rm ch}^{+-}(\omega_\nu)] . \nonumber \\ && 
\end{eqnarray}
where ${\cal D}_{\rm ch}^{pp'}(\omega_\nu)=\langle \theta_{pN}(\omega_\nu)
\theta_{p'N}(-\omega_\nu)\rangle$ [see Eq.~(\ref{Stheta})].

The conductivity is defined by $\sigma(\omega)=i/(\omega+i0^+)\Pi^{\rm
ret} (\omega)$ where the retarded correlation function $\Pi^{\rm
ret}(\omega)$ is 
obtained from $\Pi(\omega_\nu)$ by analytical continuation to real frequency
$i\omega_\nu \to \omega+i0^+$. Using the expression of $c_{pp'}$ (Appendix
\ref{app:chi}), we obtain the dissipative part of  the conductivity 
\begin{equation}
{\rm Re}[\sigma(\omega)] = \frac{\omega_p^2}{4} \Biggl [ 
\delta(\omega) \frac{3(1-\tilde\gamma^2)}{3+5\tilde\gamma^2}
+ \delta(\omega\pm \omega_0) \frac{4\tilde\gamma}{3+5\tilde\gamma^2} \Biggr].
\label{cond}
\end{equation}
where $\omega_p=\sqrt{8e^2v_F/b}$ is the plasma frequency.
Eq.~(\ref{cond}) satisfies the conductivity sum rule
\begin{equation}
\int_{-\infty}^\infty d\omega\, {\rm Re}[\sigma(\omega)] =
\frac{\omega_p^2}{4}.
\end{equation}

Thus all the spectral weight is exhausted by the collective sliding modes. 
Quasi-particle excitations above the mean-field gap do not
contribute to the (longitudinal) optical conductivity, a result well known
in SDW systems. \cite{Gruner} Because both modes contribute to the 
conductivity, the low-energy (Goldstone) mode carries only a fraction of the
total spectral weight. We obtain Dirac
peaks at $\pm \omega _0$ because we have neglected the coupling of the gapped
mode with quasi-particle excitations. Also, in a real system
(with impurities), the Goldstone mode would broaden and appear at a finite
frequency (below the quasi-particle excitation gap) due to pinning by
impurities. In the clean limit, which is appropriate in (TMTSF)$_2$X salts,
the presence of impurities does not restore any significant spectral weight to
quasi-particle excitations above the mean-field gap. \cite{Gruner} 
Therefore, the fraction of spectral weight carried by the two
modes is correctly given by Eq.~(\ref{cond}). By measuring the optical
conductivity $\sigma (\omega )$, we can therefore obtain the ratio $|\tilde
\gamma |\simeq |\gamma |$ of the amplitudes of the two SDW's. 

We have shown in Ref.~\onlinecite{Dupuis98} that the sinusoidal phase
becomes unstable against the formation of a helicoidal phase when
$|\tilde\gamma|$ reaches a critical value $\tilde\gamma_c$. By comparing the
mean-field condensation energies of the sinusoidal [Eq.~(\ref{DE})] and
helicoidal [Eq.~(\ref{DEhel}) in Sec.~\ref{sec:hel} below] phases, we find
$\tilde\gamma_c\sim 0.4$, which is in good agreement with 
the result $\tilde\gamma_c\sim 0.52$ obtained from the Ginzburg-Landau
expansion of the free energy close to the transition line. \cite{Dupuis98} 

Since $|\tilde\gamma|$ varies between 0 and 0.4 in the sinusoidal phase, the
Goldstone mode can carry between $\sim 100\% $ and $\sim 50\% $ of the total
spectral weight (Fig.~\ref{fig:sw}). When
this fraction is reduced below $\sim 50\% $ ({\it i.e.} when $|\tilde \gamma | 
\stackrel{\textstyle >}{\sim } 0.4$), a first order transition to the
helicoidal phase takes place. In the helicoidal phase, there is only one
Goldstone mode (see Sec.~\ref{sec:hel}). This mode is a pure spin-wave mode
and does not contribute to the optical spectral weight. 

Experimentally, the helicoidal phase can be stabilized by decreasing
pressure, which increases the ratio $r=g_3/g_2$
(Fig.~\ref{fig:Tc}). \cite{Dupuis98} Fig.~\ref{fig:sw1} shows
$|\tilde\gamma|$ and the low-energy spectral weight as a function of $r$ for
$H\simeq 10$ T.  In the helicoidal phase, $\tilde\gamma$ is defined by
Eq.~(\ref{gamthel}).

\section{Spin-wave modes}
\label{sec:spmode} 

For the spin-wave modes, the condition ${\rm det}{\cal D}_{\rm sp}^{-1}(\tilde
q)=0$ yields 
\begin{eqnarray}
\delta c_{++} \frac{r\zeta}{\tilde\gamma} +\delta c_{--} \frac{r\tilde
\gamma}{\zeta} +2r \delta c_{+-} && \nonumber \\ 
- (1-r^2)(\delta c_{++}\delta c_{--}-\delta c^2_{+-}) &=& 0. 
\label{sp1}
\end{eqnarray}

\subsection{Goldstone mode}

From Eq.~(\ref{sp1}) we deduce the existence of a mode with a linear dispersion
law
\begin{eqnarray}
\omega &=& v_Fq_x , \\
\phi_N(q_x,\omega) &=& \phi_{\bar N}(q_x,\omega) .
\end{eqnarray}
The spin-wave Goldstone mode corresponds to in-phase oscillations of the two 
SDW's, in agreement with the conclusion of the mean-field analysis
(Sec.~\ref{sec:MF}). We do not find any renormalization of its velocity
because we have not taken into account the coupling with the long-wavelength
spin fluctuations.\cite{Gruner}

\subsection{Gapped mode}

The other solution of (\ref{sp1}) corresponds to a gapped mode with the
dispersion law
\begin{eqnarray}
\omega^2 &=& v_F^2q_x^2 + \omega_1^2 , \nonumber \\ 
\omega_1^2 &=& \frac{12}{\tilde g_2} \frac{r}{1-r^2} 
\frac{1-\tilde\gamma^2}{\tilde\gamma^2} \Biggl | 
\frac{\tilde\Delta_N\tilde\Delta_{\bar N}}{I_NI_{\bar N}} \Biggr | .
\end{eqnarray}
The oscillations of this mode satisfy 
\begin{equation}
\phi_N(q_x,\omega) = -r\gamma \phi_{\bar N}(q_x,\omega).
\end{equation}
In the gapped mode, the oscillations of the two SDW's are out-of-phase. 
\cite{note6} As expected, the largest SDW has the smallest oscillations. 
Like the gapped sliding mode, this mode is found to lie in general above the
quasi-particle excitation gap, and is therefore expected to be strongly
damped due to coupling with quasi-particle excitations. Thus, the spin-wave
modes are similar to the sliding modes as shown in Fig.~\ref{fig:cm}.

\section{Spin-spin correlation function}
\label{sec:spsp}

In this section we calculate the transverse spin-spin correlation
function. For real order parameters $\Delta_{\alpha\sigma}^{(pN)}$, the
mean-field magnetization $\langle {\bf S}({\bf r})\rangle$ is parallel to
the $x$ axis. Transverse (to the magnetization) spin fluctuations
corresponds to fluctuations of the operator
\begin{equation}
S_y({\bf r},\tau) = \frac{i}{2} \sum_{\alpha,\sigma} \sigma O_{\alpha\sigma}
({\bf r},\tau) .
\end{equation}
The correlation function $\chi_{yy}=\langle S_yS_y\rangle$ can be expressed
as a functional derivative of the free energy with respect to an external
field that couples to the spin operator $S_y$. This standard procedure
allows to rewrite $\chi_{yy}$ as
\begin{eqnarray}
\chi_{yy}({\bf r},\tau;{\bf r}',\tau') &=& \frac{1}{4} \sum_{\alpha,\alpha',
\sigma,\sigma'} \sigma\sigma' \langle\Delta_{\alpha\sigma}({\bf r},\tau) 
\Delta^*_{\alpha'\sigma'}({\bf r}',\tau')\rangle \nonumber \\ && - 
\frac{1}{2} \delta({\bf r}-{\bf r}')\delta(\tau-\tau') \sum_{\alpha,\alpha'}
\hat g^{-1}_{\alpha\alpha'} ({\bf r}) ,
\label{chiyy}
\end{eqnarray}
where the mean value $\langle \cdots \rangle$ is taken with the effective
action $S[\Delta^*,\Delta]$. In the following, we drop the last term which
does not contribute to the spectral function ${\rm Im}\chi_{yy}^{\rm
ret}$. To first order in phase fluctuations
\begin{eqnarray}
\chi_{yy}({\bf r},\tau;{\bf r}',\tau')
&=& 4 \sum_{p,p'} \Delta_{pN} \Delta_{p'N} 
\cos({\bf Q}_{pN} \cdot {\bf r}) \nonumber \\ && \times
\cos({\bf Q}_{p'N} \cdot {\bf r}')
{\cal D}_{\rm sp}^{pp'}({\bf r},\tau;{\bf r}',\tau') ,
\end{eqnarray}
where ${\cal D}_{\rm sp}^{pp'}({\bf r},\tau;{\bf r}',\tau')= \langle \phi_{pN}
({\bf r},\tau) \phi_{p'N}({\bf r}',\tau') \rangle$. Taking the Fourier
transform, we obtain
\begin{equation}
\chi_{yy}({\bf q}+\alpha {\bf Q}_{pN},{\bf q}+\alpha' {\bf Q}_{p'N},
\omega_\nu) = \Delta_{pN} \Delta_{p'N} {\cal D}_{\rm sp}^{pp'}(\tilde q) ,
\label{chiyy1}
\end{equation}
where ${\bf q}=(q_x,q_y=0)$. 
Because of the presence of SDW's, the spin-spin correlation function is not
diagonal in momentum space. In Eq.~(\ref{chiyy1}), $\tilde q$ corresponds to
the 
momentum and energy of the spin-wave mode and tends to zero for long-wavelength
fluctuations. We therefore consider the spectral function ${\rm Im\,Tr}_{\bf
q} \chi_{yy}^{\rm ret}$, where ${\rm Tr}_{\bf q}$ is a partial trace
corresponding to a given spin-wave mode momentum ${\bf q}=(q_x,q_y=0)$:
\begin{eqnarray}
{\rm Tr}_{\bf q} \chi_{yy} &=& \sum_{\alpha,p} \chi_{yy} ({\bf q}+\alpha
{\bf Q}_{pN},{\bf q}+\alpha {\bf Q}_{pN},\omega_\nu) \nonumber \\ &=& 
2\Delta_N^2[{\cal D}_{\rm sp}^{++}(\tilde q) +\gamma^2 {\cal D}_{\rm
sp}^{--} (\tilde q)] . 
\end{eqnarray}
Using the expression of $c_{pp'}$ (Appendix \ref{app:chi}), we obtain
\begin{eqnarray}
{\rm Im}{\rm Tr}_{\bf q}\chi _{yy}^{\rm ret}&=& \frac{2\pi }{g_2^2N(0)} 
\left | \frac{\tilde \Delta _N\tilde \Delta_{\bar N}}{I_NI_{-N}} \right |
\Biggl \lbrace \frac{\delta (\omega -v_Fq_x)}{v_Fq_x}
\nonumber \\ && \times 
\Biggl \lbrack \frac{-4r}{(1-r^2)^2}+\frac{1+r^2}{(1-r^2)^2}
 \frac{\tilde \gamma ^2+\zeta^2}{|\tilde \gamma \zeta|} \Biggr \rbrack  
\nonumber \\ && +
\frac{\delta (\omega -\omega _1)}{\omega _1} \frac{3(1+r^2)}{2(1-r^2)^2}
\frac{1-\tilde \gamma ^2}{|\tilde \gamma \zeta|} \Biggr \rbrace , 
\label{Chiyy2}
\end{eqnarray}
for $\omega ,q_x>0$ and $q_y=0$.  Both spin-wave modes
contribute to the spectral function. The spectral weight carried by the
Goldstone mode diverges as $1/q_x$ as expected for a quantum antiferromagnet.
\cite{Fradkin}

Eqs.~(\ref{cond}) and (\ref{Chiyy2}) predict that all the spectral 
weight is carried by the in-phase modes, {\it i.e.} the gapped sliding mode and
the gapless spin-wave mode, whenever both SDW's have the same amplitude
($\tilde\gamma =\zeta =\pm 1$).

\section{Helicoidal phase}
\label{sec:hel}

The analysis of the helicoidal phase turns out to be much simpler than the
one of the sinusoidal phase. In the helicoidal phase, the mean-field gap
does not depend on the transverse momentum $k_y$, which significantly
simplifies the computation of the collective modes. In this section, we
shall describe the properties of the helicoidal phase, but skipping most of
the technical details of the derivation.

\subsection{Mean-field theory}

The helicoidal phase is characterized by the order parameter \cite{Dupuis98}
\begin{equation}
\Delta _{\alpha\sigma}({\bf r}) = \langle O_{\alpha\sigma}({\bf r},\tau)
\rangle = \Delta_{\alpha\sigma} e^{i\alpha {\bf Q}_{p(\alpha,\sigma)N} \cdot
{\bf r}} ,
\label{pohel}
\end{equation}
with
\begin{equation}
p(+,\uparrow) = p(-,\downarrow) = + , \,\,\,\,\, 
p(+,\downarrow) = p(-,\uparrow) = - .
\end{equation}
In the notations of the preceding sections, this corresponds to
$\Delta^{(N)}_{+\uparrow} \equiv \Delta_{+\uparrow}=\Delta^*_{-\downarrow}$,
$\Delta^{(\bar N)}_{+\downarrow} \equiv
\Delta_{+\downarrow}=\Delta^*_{-\uparrow}$, $\Delta^{(N)}_{+\downarrow}= 
\Delta^{(N)}_{-\uparrow}=0$, and  $\Delta^{(\bar N)}_{+\uparrow}= 
\Delta^{(\bar N)}_{-\downarrow}=0$. The fact that some order parameters
vanish in the helicoidal phase makes the computation of the collective modes
much simpler. The spin-density modulation is given
by\cite{Dupuis98}
\begin{eqnarray}
\langle S_x ({\bf r}) \rangle &=& 2 |\Delta_{+\uparrow}| \cos(-{\bf Q}_N \cdot
{\bf r}-\varphi_{+\uparrow} ) \nonumber \\ && 
+ 2 |\Delta_{-\uparrow}| \cos({\bf Q}_{\bar N}
\cdot {\bf r}-\varphi_{-\uparrow} ), \nonumber \\ 
\langle S_y ({\bf r}) \rangle &=& 2 |\Delta_{+\uparrow}| \sin(-{\bf Q}_N \cdot
{\bf r}-\varphi_{+\uparrow} )  \nonumber \\ && 
+ 2 |\Delta_{-\uparrow}| \sin({\bf Q}_{\bar N}
\cdot {\bf r}-\varphi_{-\uparrow} ) , \nonumber \\ 
\langle S_z ({\bf r}) \rangle &=& 0 , 
\end{eqnarray}
which corresponds to two helicoidal SDW's with opposite chiralities.
The mean-field action is still given by (\ref{action12}), but the pairing
amplitudes are given by (in the QLA) 
\begin{eqnarray} 
\tilde\Delta_{\alpha\sigma}({\bf k},{\bf k}') &=& \delta_{{\bf k}',{\bf
k}-\alpha {\bf Q}_0} \tilde\Delta_{\alpha\sigma}(k_y) , \nonumber \\ 
\tilde\Delta_{\alpha\sigma}(k_y) &=& \tilde\Delta_{\alpha\sigma} 
e^{i\alpha p(\alpha,\sigma)N(k_yb-\pi/2)} , \nonumber \\ 
\tilde\Delta_{\alpha\sigma} &=& I_{p(\alpha,\sigma)N} (g_2
\Delta_{\alpha\sigma} + g_3 \Delta_{\bar\alpha\sigma} ) . 
\end{eqnarray}

The ground-state energy reads
\begin{eqnarray}
\Delta E &=& \sum_\alpha \frac{1}{I_{p(\alpha,\uparrow)N}} \Delta^*
_{\alpha\uparrow} \tilde \Delta_{\alpha\uparrow} \nonumber \\ && 
-\frac{N(0)}{2} \sum_\alpha
\biggl [ \frac{|\tilde\Delta_{\alpha\uparrow}|^2}{2}
+ |\tilde\Delta_{\alpha\uparrow}|^2 \ln
\frac{2E_0}{|\tilde\Delta_{\alpha\uparrow}|} \biggr ] \nonumber \\ &=& 
- \frac{1}{4} N(0) \sum_\alpha |\tilde\Delta_{\alpha\uparrow}|^2 ,
\label{DEhel}
\end{eqnarray}
where the last line is obtained using the gap equation $\partial \Delta
E/\partial \tilde\Delta^*_{\alpha\uparrow}=0$. From the latter, we also
deduce (writing $\tilde\Delta_{\alpha\sigma}= |\tilde\Delta_{\alpha\sigma}|
e^{i\tilde \varphi_{\alpha\sigma}}$) 
\begin{equation}
\tilde \varphi_{+\uparrow}-\tilde \varphi_{-\uparrow} = 
\left \lbrace 
\begin{array}{lr}
0   & {\rm if} \,\,\,  I_NI_{\bar N} > 0, \\
\pi & {\rm if} \,\,\,  I_NI_{\bar N} < 0, 
\end{array} \right.
\end{equation}
and
\begin{equation}
\varphi_{+\uparrow}=\varphi_{-\uparrow} .
\end{equation}
Without loss of generality, we can choose the order parameters real, and
introduce 
\begin{equation}
\gamma = \frac{\Delta_{-\uparrow}}{\Delta_{+\uparrow}} , \,\,\, 
\tilde\gamma =  \frac{\tilde\Delta_{-\uparrow}}{\tilde\Delta_{+\uparrow}} .
\label{gamthel}
\end{equation}
$\gamma$ and $\tilde\gamma$ satisfy Eqs.~(\ref{gam2t}). 

The mean-field propagators are then 
\begin{eqnarray}
G_{\alpha\sigma}({\bf k},\omega) &=& \frac{-i\omega-\epsilon_{\alpha\sigma}
(k_x)}{\omega^2+\epsilon^2_{\alpha\sigma}(k_x)+|\tilde\Delta_{\alpha\sigma}|^2}
, \nonumber \\ 
F_{\alpha\sigma}({\bf k},\omega) &=& \frac{\tilde\Delta_{\alpha\sigma}
e^{i\alpha p(\alpha,\sigma)N(k_yb-\pi/2)}}{\omega^2+\epsilon^2
_{\alpha\sigma}(k_x) +|\tilde\Delta _{\alpha\sigma}|^2} .
\end{eqnarray}
The excitation energy $E_{\alpha\sigma}({\bf k}) =
[\epsilon^2_{\alpha\sigma}(k_x)+|\tilde\Delta_{\alpha\sigma}|^2]^{1/2}$ does
not depend on $k_y$. The spectrum is shown in Fig.~\ref{fig:spec}. Note that
all the branches are gapped, due to the presence of two SDW's. In contrast, in
presence of a single helicoidal SDW, some branches would remain gapless. 

As in the sinusoidal phase, we can compute the mean-field susceptibilities. 
They are given in Appendix \ref{app:hel}.

\subsection{Collective modes}

Considering phase fluctuations only, we have
\begin{eqnarray}
\eta _{\alpha\sigma}({\bf r},\tau) &=& 
\Delta _{\alpha\sigma} e^{i\alpha {\bf Q}_{p(\alpha,\sigma)N} \cdot {\bf r}
+i \varphi_{\alpha\sigma} ({\bf r},\tau) } -\Delta_{\alpha\sigma} ({\bf
r},\tau) \nonumber \\ 
&\simeq & i\Delta_{\alpha\sigma}({\bf r}) \varphi_{\alpha\sigma}({\bf
r},\tau) 
\end{eqnarray}
to lowest order in phase fluctuations. Following Sec.~\ref{sec:Seff}, we
then obtain the effective action
\begin{equation}
S[\varphi] = \frac{1}{2} \sum_{\tilde q} (\varphi_{+\uparrow}(-\tilde q) ,
\, \varphi_{-\uparrow}(-\tilde q)) {\cal D}^{-1}(\tilde q) 
\left (
\begin{array}{r} 
\varphi_{+\uparrow}(\tilde q) \\
\varphi_{-\uparrow}(\tilde q) 
\end{array}
\right ) , 
\end{equation}
\bleq
\begin{equation}
{\cal D}^{-1}(\tilde q) = 2g_2\Delta^2_{+\uparrow} 
\left (
\begin{array}{lr}
1-c_{++} -r^2c_{--} & \gamma r(1-c_{++} -c_{--}) \\
\gamma r(1-c_{++} -c_{--}) & \gamma^2 (1-c_{--} -r^2c_{++})
\end{array} 
\right ) .
\end{equation} 
\eleq

The dispersion of the collective modes is obtained from ${\rm det}{\cal
D}^{-1} (\tilde q) =0$. We find a Goldstone mode satisfying
\begin{equation}
\omega =v_Fq_x, \,\,\,
\varphi_{+\uparrow}=\varphi_{-\uparrow}. 
\end{equation}
This mode corresponds to a uniform spin rotation. Equivalently, it can also
be seen as a translation of the two SDW's in 
opposite directions.\cite{note4} Thus it combines characteristics of the two
Goldstone  modes of the sinusoidal phase.

There is also a gapped mode with
dispersion law $\omega^2=v_F^2q_x^2+\omega_2^2$, where
\begin{equation}
\omega_2^2= \frac{16}{\tilde g_2} \frac{r}{1-r^2} \Biggl |
\frac{\tilde\Delta_{+\uparrow} \tilde\Delta_{-\uparrow}}{I_NI_{\bar N}}
\Biggr | . 
\label{om2}
\end{equation}
The oscillations satisfy
\begin{equation}
\frac{\varphi_{+\uparrow}}{\varphi_{-\uparrow}} = 
\frac{\gamma^2(1-r^2)^2+2r\gamma(\gamma+r)^2+2r\gamma(1+\gamma r)^2}{
\gamma^2(1-r^2)^2-2(\gamma+r)^2-2r^2 (1+\gamma r)^2} . 
\end{equation} 
When the two SDW's have the same amplitude ($\gamma=1$),
$\varphi_{+\uparrow} /\varphi_{-\uparrow}=-1$. As for the sinusoidal phase,
the gapped mode is found to lie above the quasi-particle excitation gap. 

The similarities with phase modes in two-band or bilayer superconductors are
more pronounced in the helicoidal phase than in the sinusoidal phase. In
fact, Eq.~(\ref{om2}) gives exactly the gap of the phase modes in these
superconducting systems. In the helicoidal phase, the order parameter
$\Delta_{\alpha\sigma}({\bf r})$ [Eq.~(\ref{pohel})] and the effective
potential $\tilde\Delta_{\alpha\sigma}({\bf r})=g_2
\Delta_{\alpha\sigma}({\bf r})+g_3e^{i\alpha 4k_Fx} 
\Delta_{\bar\alpha\sigma}({\bf r})$ have only one component $\sim e^{i\alpha
{\bf Q}_{p(\alpha,\sigma)N}\cdot {\bf r}}$. Thus, the helicoidal phase is
more BCS-like than the sinusoidal phase. This explains why the
quasi-particle excitation spectrum does not depend on $k_y$, and also the
absence of a $\tilde\gamma$-dependent factor in the energy of the gapped
mode [Eq.~(\ref{om2})]. We have shown in Sec.~\ref{sec:opt} that in the
sinusoidal phase only the largest gap remains tied to the Fermi level in
presence of condensate fluctuations. This yields a  modification of the
condensation energy, which is at the origin of the $\tilde\gamma$-dependent
factors in the expressions of $\omega_0$ and $\omega_1$.

\subsection{Spectral functions}

In the helicoidal phase, it is not clear whether the collective modes should
be seen as sliding or spin-wave modes.\cite{note4} In this section, we
compute the spectral functions ${\rm Re}[\sigma(\omega)]$ and ${\rm Im}{\rm
Tr}_{\bf q} \chi^{\rm ret}$ to answer this question. 

\subsubsection{Optical conductivity}

The charge and current operators are given by
\begin{eqnarray}
\rho_{\rm DW} ({\bf r},\tau) &=& \frac{1}{2\pi b} \partial_x \Bigl (
\tilde\varphi_{+\uparrow}({\bf r},\tau) - \tilde\varphi_{-\uparrow}({\bf
r},\tau) \Bigr ) ,  \\ 
j_{\rm DW}({\bf r},\tau) &=& - \frac{ie}{2\pi b} \partial_\tau  \Bigl (
\tilde\varphi_{+\uparrow}({\bf r},\tau) - \tilde\varphi_{-\uparrow}({\bf
r},\tau) \Bigr ) ,
\label{cur2}
\end{eqnarray}
where
\begin{equation}
\tilde\varphi_{\alpha\sigma}({\bf r},\tau) = 
\frac{g_2\Delta_{\alpha\sigma}\varphi_{\alpha\sigma}({\bf r},\tau) + 
g_3 \Delta_{\bar\alpha\sigma}\varphi_{\bar\alpha\sigma}({\bf r},\tau)}{g_2
\Delta_{\alpha\sigma}+g_3\Delta_{\bar\alpha\sigma}} 
\end{equation} 
is the fluctuating phase of the effective potential
$\tilde\Delta_{\alpha\sigma}$.
From Eq.~(\ref{cur2}) we can calculate the current-current correlation
function, which yields the conductivity
\begin{equation}
{\rm Re}[\sigma(\omega)] = \frac{\omega_p^2}{8} \delta (\omega\pm\omega_2) .
\end{equation}
The Goldstone mode does not contribute to the conductivity and is therefore
a pure spin-wave mode (in the limit ${\bf q}\to 0$). This result could have
been anticipated since the condition
$\varphi_{+\uparrow}=\varphi_{-\uparrow}$  implies $\tilde\varphi_{+\uparrow}=
\tilde\varphi_{-\uparrow}$ and the vanishing of the
current (\ref{cur2}). Thus all the spectral weight is pushed up above the
quasi-particle excitation gap at frequencies of the order of
$\omega_2$. This implies that there is no low-energy collective mode
corresponding to a uniform sliding of the condensate. Since non-linear
conduction in SDW systems results from the depinning of such a mode (from
the impurity potential) above a threshold electric field, we conclude to the
absence of non-linear conductivity in the helicoidal phase. 

In Ref.~\onlinecite{Dupuis98}, we have shown that the helicoidal phase is
characterized by a vanishing QHE ($\sigma_{xy}=0$) and a kinetic
magneto-electric effect. By studying the collective modes, we have found a
third possibility for the experimental detection of the helicoidal phase,
namely the absence of non-linear conduction.

\subsubsection{Spin-spin correlation function}

We consider the spin-spin correlation function 
\begin{equation}
\chi_{\mu\nu}({\bf r},\tau;{\bf r}',\tau') = \langle S_\mu({\bf r},\tau) 
S_\nu({\bf r}',\tau') \rangle .
\end{equation}
As in Sec.~\ref{sec:spsp}, we shall compute the partial trace ${\rm
Tr}_{\bf q} \chi_{\mu\nu}$, where the momentum $\bf q$ of the collective
mode is held fixed. [Again, we drop the last term of (\ref{chiyy}).] We find
\begin{eqnarray}
{\rm Tr}_{\bf q} \chi_{\mu\mu} &=& \frac{1}{2}
\sum_\alpha \Delta^2_{\alpha\uparrow} {\cal D}_{\alpha\alpha} (\tilde q) ,
\nonumber \\ 
{\rm Tr}_{\bf q} \chi_{xy} &=& 0 .
\end{eqnarray}
This yields the spectral function
\bleq
\begin{eqnarray}
{\rm Im} {\rm Tr}_{\bf q} \chi^{\rm ret}_{\mu\mu} &=& \frac{\pi}{2g_2^2N(0)} 
\Biggl | \frac{\tilde\Delta_{+\uparrow}\tilde\Delta_{-\uparrow}}{I_N I_{\bar
N}} \Biggr | \Biggl \lbrace 
\frac{\delta(\omega-v_Fq_x)}{v_Fq_x} 
\Biggl [ -\frac{4r}{(1-r^2)^2} +
\frac{1+r^2}{(1-r^2)^2} \frac{\tilde\gamma^2+\zeta^2}{\tilde\gamma\zeta} 
\Biggr ] \nonumber \\ && + 
\frac{\delta(\omega-\omega_2)}{\omega_2}  \Biggl [ \frac{4r}{(1-r^2)^2} +
\frac{1+r^2}{(1-r^2)^2} \frac{\tilde\gamma^2+\zeta^2}{\tilde\gamma\zeta} 
\Biggr ] \Biggr \rbrace \nonumber \\ && 
\end{eqnarray}
\eleq
for $\omega,q_x>0$ and $q_x\to 0$. Both modes contribute to the spectral
function.  Although the Goldstone mode
is a pure spin-wave mode, the gapped mode has characteristics of both a spin
wave and a sliding mode, as can be seen from the spectral function.

\section{Conclusion}

We have studied the long-wavelength collective modes in the FISDW phases of
quasi-1D conductors, focusing on phases that exhibit a sign reversal of the
QHE (Ribault anomaly). We have recently proposed that two SDW's, with wave
vectors ${\bf Q}_N=(2k_F+NG,Q_y)$ and ${\bf Q}_{\bar N}=(2k_F-NG,-Q_y)$, 
coexist in the Ribault phase, as a result of umklapp scattering. When the
latter is strong enough, the two SDW's become circularly polarized (helicoidal
SDW's). The presence of two SDW's gives rise to a rich structure of collective
modes, which strongly depends on the polarization (linear or circular) of the
SDW's. 

Regarding the sliding modes, we find that the out-of-phase oscillations are
gapless in the long-wavelength limit. The fact that this Goldstone mode 
corresponds to out-of-phase (and not in-phase) oscillations is related to the
pinning by the lattice (due to umklapp processes) that would occur for a
single commensurate SDW. The other sliding mode is gapped and corresponds to
in-phase oscillations. In Bechgaard salts, this mode is expected to lie above
the quasi-particle excitation gap and should therefore be strongly damped due 
to the coupling with the quasi-particle excitations. In the helicoidal phase,
there is no low-energy sliding mode, since the Goldstone mode is a pure 
spin-wave mode. [For a helicoidal SDW, one cannot distinguish
between a uniform spin rotation and a global translation, so that we cannot 
classify the modes in sliding and spin-wave modes.] 

The dissipative part of the conductivity, ${\rm Re}[\sigma(\omega)]$, 
exhibits two peaks: a
low-energy peak corresponding to the Goldstone mode, and a  (broader) peak 
at high energy due to the incoherent gapped mode. The low-energy spectral
weight is directly related to the ratio of the amplitudes of the two SDW's 
that coexist in the Ribault phase. When the umklapp scattering strength ($g_3$)
increases (experimentally this corresponds to a pressure decrease), spectral
weight is transfered from the low-energy peak to high energies. Above a
critical value of $g_3$, the sinusoidal phase becomes unstable with respect to
the formation of a helicoidal phase. At the transition, the low-energy spectral
weight suddenly drops to zero (Fig.~\ref{fig:sw}), since there is no low-energy
optical spectral weight in the helicoidal phase. Thus, the formation of the 
helicoidal phase can be detected by measuring the low-energy optical spectral
weight. We also note that the absence of a low-energy sliding mode means that
there is no infinite Fr\"ohlich conductivity\cite{Frohlich54} in this 
helicoidal phase, which is therefore a true insulating phase even in an ideal 
system ({\it i.e.} with no impurities). In a real system (with impurities), 
this implies the absence of non-linear dc conductivity. 

The spin-wave modes exhibit a similar structure. The in-phase oscillations of 
the two SDW's are gapless (Goldstone mode), while the out-of-phase oscillations
are gapped. In the helicoidal phase, the Goldstone mode is a pure spin-wave
mode, but the gapped mode contributes both to the conductivity and to the 
transverse spin-spin correlation function. 

As discussed in the Introduction, these conclusions rely on a simple Fermi
surface (Fig.~\ref{fig:FS}a), which does not necessarily provide a good 
approximation to the actual Fermi surface of the Bechgaard salts. They 
should therefore be taken with caution 
regarding their relevance to the organic conductors of the Bechgaard salts
family. However, we have derived a number of experimental consequences that
should allow to test our theory. In the sinusoidal phase, we predict a possible
reentrance of the phase $N=0$ within the cascade.\cite{Dupuis98} The low-energy
peak in the optical conductivity ${\rm Re}[\sigma(\omega)]$ carries only a 
fraction of the total spectral weight $\omega_p^2/4$. This fraction should 
decrease with pressure. At low pressure, the sinusoidal phase may become 
helicoidal. The helicoidal phase is characterized by a vanishing QHE, 
\cite{Dupuis98} a kinetic magneto-electric effect, \cite{Dupuis98} and the 
absence of low-energy spectral weight in the optical conductivity as well as
the absence of non-linear dc conductivity. In the alternative scenario proposed
by Zanchi and Montambaux,\cite{Zanchi96} the Ribault phase does not exhibit 
any special features compared to the positive phases, apart from the unusual 
behavior of the magneto-roton modes,\cite{Lederer98} but these modes have not
been observed yet. 

We expect our conclusions regarding the structure of the collective modes and 
the associated spectral functions to hold for generic two-SDW systems and not
only for the FISDW phases that exhibit the Ribault anomaly. It is clear that 
the existence of four long-wavelength collective modes (two spin-wave and two
sliding modes) results from the presence of a second SDW, which doubles the 
number of degrees of freedom. Two of these modes should be gapless as expected
from Goldstone theorem in a system where two continuous symmetries are 
spontaneously broken: the translation symmetry in real space and the rotation 
symmetry in spin space. 

This belief is supported by the striking analogy with collective modes in
other systems like two-band, bilayer, and $d+id'$ superconductors, 
\cite{Legett66,Hwang98,Balatsky99} and, to a lesser extent, plasmon modes
in semiconductor double-well structures.\cite{Dassarma81} While phase modes
are in general difficult to observe in superconductors, since they do not
directly couple to external probes (see however Ref.~\onlinecite{Balatsky99}),
collective modes in SDW systems directly show up in response functions like 
for instance the dc and optical conductivities.  

Finally, we point out that we expect a similar structure of collective modes 
in the FISDW phases of the compound (TMTSF)$_2$ClO$_4$. Due to anion
ordering, the unit cell contains two sites, which leads to two electronic bands
at the Fermi level. This doubles the number of degrees of freedom and therefore
the number of collective modes (without considering the possible formation of
a second SDW due to umklapp scattering). As for phonon modes in a crystal with
two molecules per unit sites, we expect an acoustic (Goldstone) mode and an
optical mode.

\section*{Acknowledgment} 

This work was partially supported by the NSF under Grant DMR9815094 and by
the Packard Foundation. Laboratoire de Physique des Solides is associ\'e au
CNRS.

\bleq

\appendix

\section{Mean-field susceptibilities $\bar\chi$ and $\tilde\chi$}
\label{app:chi}

Using the expression (\ref{propa1}) of the mean-field propagators $G$ and
$F$, we obtain
\begin{eqnarray}
\bar\chi_{\alpha\sigma,\alpha\sigma}({\bf q}+\alpha {\bf Q}_{pN},{\bf q}'
+\alpha {\bf Q}_{p'N},\omega_\nu) &=& -\delta_{{\bf q},{\bf q}'}
\frac{T}{L_xL_y} \sum_{{\bf k},\omega} \sum_{n=-\infty}^\infty 
G_{\alpha\sigma}({\bf k},\omega) 
G_{\bar\alpha\bar\sigma}({\bf k}-\alpha {\bf Q}_0-{\bf q}+\alpha
nG,\omega-\omega_\nu) \nonumber \\ && \times
I_{pN+n}(\pi/b+q_y) I_{p'N+n}(\pi/b+q_y)
e^{-i\alpha(p-p')N(k_yb-q_yb/2-\pi/2)} , \\
\bar\chi_{\alpha\sigma,\bar\alpha\bar\sigma}({\bf q}+\alpha {\bf
Q}_{pN},{\bf q}'-\alpha {\bf Q}_{p'N},\omega_\nu) &=& -\delta_{{\bf q},{\bf
q}'} \frac{T}{L_xL_y} \sum_{{\bf k},\omega} \sum_{n=-\infty}^\infty 
F_{\alpha\sigma}({\bf k},\omega) 
F_{\alpha\sigma}({\bf k}-{\bf q}+\alpha nG,\omega-\omega_\nu) 
\nonumber \\ && \times 
I_{pN+n}(\pi/b+q_y) I_{p'N-n}(\pi/b+q_y)
e^{-i\alpha(p+p')N(k_yb-q_yb/2-\pi/2)} .
\end{eqnarray}
Here we have assumed that $|q_x|,|q_x'|\ll G$, which is the case of interest
when studying the low-energy fluctuations around the mean-field solution. 

In the QLA, we retain only the term $n=0$ in the above equations, since the
other terms are strongly suppressed when $\omega_c \gg T,|\tilde\Delta
_{pN}|$ in the limit of long-wavelength fluctuations.\cite{Poilblanc87a} 
Restricting ourselves to $q_y=0$, we then obtain
\begin{equation}
\bar\chi_{\alpha\sigma,\alpha\sigma}({\bf q}+\alpha {\bf Q}_{pN},{\bf q}
+\alpha {\bf Q}_{p'N},\omega_\nu) = -
\frac{T}{L_xL_y} \sum_{{\bf k},\omega} 
G_{\alpha\sigma}({\bf k},\omega) 
G_{\bar\alpha\bar\sigma}({\bf k}-\alpha {\bf Q}_0-{\bf q},
\omega-\omega_\nu) I_{pN} I_{p'N}
e^{-i\alpha(p-p')N(k_yb-\pi/2)} ,
\end{equation}
\begin{equation}
\bar\chi_{\alpha\sigma,\bar\alpha\bar\sigma}({\bf q}+\alpha {\bf
Q}_{pN},{\bf q}-\alpha {\bf Q}_{p'N},\omega_\nu) = -
\frac{T}{L_xL_y} \sum_{{\bf k},\omega}
F_{\alpha\sigma}({\bf k},\omega) 
F_{\alpha\sigma}({\bf k}-{\bf q},\omega-\omega_\nu) 
I_{pN} I_{p'N}
e^{-i\alpha(p+p')N(k_yb-\pi/2)} .
\end{equation}
Performing the sum over $k_x$, we obtain at $T=0$ and in the limit
$|v_Fq_x|,|\omega_\nu|\ll |\tilde\Delta_{\alpha\sigma}(k_y)|$
\begin{eqnarray}
\bar\chi_{\alpha\sigma,\alpha\sigma}({\bf q}+\alpha {\bf Q}_{pN},{\bf q}
+\alpha {\bf Q}_{p'N},\omega_\nu) &=& I_{pN} I_{p'N} \frac{N(0)}{2N_\perp}
\sum_{k_y} e^{-i\alpha(p-p')N(k_yb-\pi/2)} \Biggl [ 
\ln \frac{2E_0}{|\tilde\Delta_{\sigma\alpha}(k_y)|} - \frac{1}{2} 
- \frac{\omega_\nu^2+v_F^2q_x^2}{6|\tilde\Delta_{\alpha\sigma}(k_y)|^2} 
\Biggr ] , 
\label{C5a} \\ 
\bar\chi_{\alpha\sigma,\bar\alpha\bar\sigma}({\bf q}+\alpha {\bf
Q}_{pN},{\bf q}-\alpha {\bf Q}_{p'N},\omega_\nu) &=& 
I_{pN} I_{p'N} \frac{N(0)}{2N_\perp}
\sum_{k_y} e^{-i\alpha(p+p')N(k_yb-\pi/2)} \Biggl [ - \frac{1}{2} 
+ \frac{v_F^2q_x^2+\omega_\nu^2}{12|\tilde\Delta_{\alpha\sigma}(k_y)|^2} 
\Biggr ] .
\label{C5b}
\end{eqnarray}
Eqs.~(\ref{C5a}) and (\ref{C5b}) are obtained by expanding to first order in
$\omega_\nu^2/|\tilde\Delta_{\alpha\sigma}(k_y)|^2$ and
$v_F^2q_x^2/|\tilde\Delta_{\alpha\sigma}(k_y)|^2$. This calculation is
standard when evaluating the long-wavelength collective modes of a SDW
system and can be found for instance in Ref.~\onlinecite{Poilblanc87a}.  
Eqs.~(\ref{chi1}) are then obtained by summing over $k_y$. 

It is useful to introduce the notations 
\begin{equation}
c_{pp'} = g_2 [\bar\chi_{+\uparrow,+\uparrow}({\bf q}+ {\bf
Q}_{pN},{\bf q} + {\bf Q}_{p'N},\omega_\nu) -
\bar\chi_{+\uparrow,-\downarrow}({\bf q}+ {\bf Q}_{pN},{\bf q} 
- {\bf Q}_{p'N},\omega_\nu)]
=  \bar c_{pp'} + \delta c_{pp'} ,
\label{chi1a}
\end{equation}
where $\bar c_{pp'}=c_{pp'} | _{{\bf q}=\omega_\nu =0}$ and $\delta c_{pp'}$
are deduced from Eqs.~(\ref{chi1}).

The gap equation can be written as a function of the static
($\omega_\nu=0$) susceptibilities $\bar\chi$ (or $\bar c$). From 
$\Delta _{\alpha\sigma}({\bf r}) = \langle O_{\alpha\sigma} ({\bf r}) \rangle
= T\sum_\omega F_{\alpha\sigma}({\bf r},{\bf r},\omega)$, we deduce  
\begin{equation}
\Delta_{pN} = I_{pN} \frac{T}{L_xL_y} \sum_{{\bf k},\omega,p'} 
\frac{e^{-i (p-p')N(k_yb-\pi/2)}}{\omega^2+\epsilon^2_{+\uparrow}
(k_x) +|\tilde\Delta_{+\uparrow}(k_y)|^2} \tilde\Delta_{p'N} .
\end{equation}
This can be rewritten in terms of the mean-field propagators
\begin{eqnarray}
\Delta_{pN} &=& - I_{pN} \frac{T}{L_xL_y}  \sum_{{\bf
k},\omega,p'} e^{-i (p-p')N(k_yb-\pi/2)}
[ G_{+\uparrow}({\bf k},\omega)  G_{-\downarrow}({\bf k}-{\bf Q}_0,\omega)
-|F_{+\uparrow}({\bf k},\omega)|^2  ] \tilde\Delta_{p'N} 
\nonumber \\ &=& 
\sum_{p'} \frac{1}{I_{p'N}} \Bigl [ 
\bar\chi_{+\uparrow,+\uparrow}({\bf Q}_{pN}, {\bf
Q}_{p'N},\omega_\nu=0) - \bar\chi_{+\uparrow,-\downarrow} ({\bf
Q}_{pN},- {\bf Q}_{p'N},\omega_\nu=0) \Bigr ] \tilde \Delta_{p'N} . 
\end{eqnarray}
Using the relation between $\Delta_{pN}$ and $\tilde
\Delta_{pN}$, we obtain
\begin{equation}
\frac{\tilde\Delta_N}{(1-r^2)I_N} - \frac{r\tilde\Delta_{\bar N}}{(1-r^2)
I_{\bar N}} = \frac{\bar c_{++}\tilde\Delta_N}{I_N} , \,\,\,\,\,\,\,\, 
\frac{\tilde\Delta_{\bar N}}{(1-r^2)I_{\bar N}} - \frac{r\tilde\Delta_N}{(
1-r^2) I_N} = \frac{\bar c_{--}\tilde\Delta_{\bar N}}{I_{\bar N}} .
\label{C10}
\end{equation}
From (\ref{C10}), we deduce the relations 
\begin{equation}
\bar c_{++} = \frac{1}{1+\gamma r} , \,\,\,
\bar c_{--} = \frac{\gamma}{\gamma +r} ,
\label{cbar}
\end{equation}
which gives, by eliminating $\gamma$,
\begin{equation}
(1-\bar c_{++})(1-\bar c_{--})-r^2 \bar c_{++}c_{--} = 0.
\label{gap3} 
\end{equation}
Eq.~ (\ref{gap3}) is nothing but the gap equation rewritten in terms of the
static mean-field susceptibilities $\bar c$. 

In the study of the collective modes, the natural quantity to consider is
not $\bar\chi$ but the susceptibility $\tilde\chi$ defined by
\begin{equation}
\tilde\chi_{\alpha\sigma,\alpha'\sigma'} ({\bf r},\tau;{\bf r}',\tau') =
\langle \tilde O_{\alpha\sigma} 
({\bf r},\tau)  \tilde O^*_{\alpha'\sigma'} ({\bf r}',\tau') \rangle -   
\langle \tilde O_{\alpha\sigma} ({\bf r},\tau) \rangle \langle \tilde
O^*_{\alpha'\sigma'} ({\bf r}',\tau') \rangle ,
\label{chiT}
\end{equation}
where $\tilde O_\sigma ({\bf r},\tau) = \hat g({\bf r}) O_\sigma 
({\bf r},\tau)$. $\tilde \chi$ is related to $\bar\chi$ by 
\begin{eqnarray}
\tilde\chi_{\alpha\sigma,\alpha'\sigma'}({\bf q},{\bf q}',\omega_\nu) &=& 
g_2^2 \bar\chi_{\alpha\sigma,\alpha'\sigma'}({\bf q},{\bf q}',\omega_\nu)
+ g_3^2 
\bar\chi_{\bar\alpha\sigma,\bar\alpha'\sigma'}({\bf q}-\alpha 4k_F, {\bf
q}'-\alpha' 4k_F,\omega_\nu) \nonumber \\ && + g_2 g_3 [
\bar\chi_{\bar\alpha\sigma,\alpha'\sigma'}({\bf q}-\alpha 4k_F, {\bf
q}',\omega_\nu)  + 
\bar\chi_{\alpha\sigma,\bar\alpha'\sigma'}({\bf q}, {\bf
q}'-\alpha' 4k_F,\omega_\nu) ]
\end{eqnarray}
and satisfies
\begin{eqnarray}
\tilde\chi_{\alpha\uparrow,\alpha'\uparrow}({\bf q}+\alpha {\bf Q}_{pN}, 
{\bf q}+\alpha' {\bf Q}_{p'N},\omega_\nu) -
\tilde\chi_{\alpha\uparrow,\bar\alpha'\downarrow}({\bf q}+\alpha {\bf Q}_{pN}, 
{\bf q}-\alpha' {\bf Q}_{p'N},\omega_\nu) &=&
g_2 \delta_{\alpha,\alpha'} ( c_{pp'} + r^2 c_{\bar p \bar p'}) 
\nonumber \\ && 
+g_3  \delta_{\alpha,\bar\alpha'} (c_{\bar pp'} + c_{p \bar p'}) .
\label{chiT1}
\end{eqnarray}
Eq.~(\ref{chiT1}) is used in Sec.~\ref{sec:Seff}.

\section{Charge operator $\rho_{\rm DW}$}
\label{app:rho}

Using 
\begin{equation}
{\cal G}_\alpha^{\rm MF}({\bf r},\tau;{\bf r}',\tau') = 
\left (
\begin{array}{lr}
G_{\alpha\uparrow} ({\bf r},\tau;{\bf r}',\tau') & 
F_{\alpha\uparrow} ({\bf r},\tau;{\bf r}',\tau') \\
F_{\bar\alpha\downarrow} ({\bf r},\tau;{\bf r}',\tau') & 
G_{\bar\alpha\downarrow} ({\bf r},\tau;{\bf r}',\tau')
\end{array}
\right ) , 
\end{equation}
where $G$ and $F$ are the mean-field propagators (see
Sec.~\ref{sec:MFpro}), we rewrite Eq.~(\ref{op3}) as
\begin{eqnarray}
\rho_{\rm DW} ({\bf r},\tau)&=& - \sum_\alpha \int d^2r'\,d\tau'\, \Bigl [ 
G_{\alpha\uparrow} ({\bf r},\tau;{\bf r}',\tau')
\tilde\eta_{\alpha\uparrow}({\bf r}',\tau') 
F_{\bar\alpha\downarrow} ({\bf r}',\tau';{\bf r},\tau) + 
F_{\alpha\uparrow} ({\bf r},\tau;{\bf r}',\tau')
\tilde\eta^*_{\alpha\uparrow}({\bf r}',\tau') 
G_{\alpha\uparrow} ({\bf r}',\tau';{\bf r},\tau) \nonumber \\ && + 
F_{\bar\alpha\downarrow} ({\bf r},\tau;{\bf r}',\tau')
\tilde\eta_{\alpha\uparrow}({\bf r}',\tau') 
G_{\bar\alpha\downarrow} ({\bf r}',\tau';{\bf r},\tau)  +
G_{\bar\alpha\downarrow} ({\bf r},\tau;{\bf r}',\tau')
\tilde\eta^*_{\alpha\uparrow}({\bf r}',\tau') 
F_{\alpha\uparrow} ({\bf r}',\tau';{\bf r},\tau) \Bigl ] . \nonumber \\ 
\label{op4}
\end{eqnarray}

If we consider phase fluctuations only, $\eta_{\alpha\sigma}$ is given by
(\ref{etaT}), which gives using (\ref{etaT1})
\begin{equation}
\tilde\eta_{\alpha\sigma}({\bf r},\tau) = i \sum_p e^{i\alpha {\bf Q}_{pN}
\cdot {\bf r}} [g_2\Delta_{pN}\varphi^{(pN)}_{\alpha\sigma} ({\bf r},\tau ) 
+g_3 \Delta_{\bar pN}\varphi^{(\bar pN)}_{\bar\alpha\sigma} ({\bf r},\tau )] . 
\end{equation}
Defining the fluctuating phases $\tilde\varphi^{(pN)}_{\alpha\sigma}$ of the
effective potential $\tilde \Delta _{\alpha\sigma}$ by
\begin{equation}
\tilde \Delta _{\alpha\sigma}({\bf r},\tau) = \sum_p
\frac{\tilde\Delta_{pN}}{I_{pN}} e^{i\alpha {\bf Q}_{pN} \cdot {\bf r} + 
i \tilde\varphi^{(pN)}_{\alpha\sigma} ({\bf r},\tau) } , 
\end{equation}
we have
\begin{equation}
\tilde\eta_{\alpha\sigma}({\bf r},\tau) = i \sum_p 
\frac{\tilde\Delta_{pN}}{I_{pN}} e^{i\alpha {\bf Q}_{pN} \cdot {\bf r}} 
\tilde\varphi^{(pN)}_{\alpha\sigma} ({\bf r},\tau) . 
\label{op5}
\end{equation}
The relation between $\tilde\varphi^{(pN)}_{\alpha\sigma}$ and
$\varphi^{(pN)}_{\alpha\sigma}$ is given by (\ref{phiT3}) to lowest order in
phase fluctuations. 

From Eqs.~(\ref{op4}) and (\ref{op5}), we deduce 
\begin{eqnarray}
\rho_{\rm DW} (\tilde q) &=& -i\frac{T}{L_xL_y} \sum_p \frac{\tilde\Delta_{pN}
}{I_{pN}} \sum_{\alpha,\tilde q'} \tilde\varphi^{(pN)}_{\alpha\uparrow}
(\tilde q') \int d^2r\,d\tau\, \int d^2r'\,d\tau'\, e^{-i({\bf q}\cdot {\bf
r}-\omega_\nu \tau) + i({\bf q}'\cdot {\bf r}'-\omega_\nu' \tau') }
\nonumber \\ && \times 
\Bigl \lbrace e^{i\alpha {\bf Q}_{pN} \cdot {\bf r}'} 
[G_{\alpha\uparrow}({\bf r},\tau;{\bf r}',\tau') 
F_{\bar\alpha\downarrow} ({\bf r}',\tau';{\bf r},\tau) + 
F_{\bar\alpha\downarrow} ({\bf r},\tau;{\bf r}',\tau')
G_{\bar\alpha\downarrow} ({\bf r}',\tau';{\bf r},\tau) ] \nonumber \\ && -
e^{-i\alpha {\bf Q}_{pN} \cdot {\bf r}'} 
[ F_{\alpha\uparrow}({\bf r},\tau;{\bf r}',\tau')
G_{\alpha\uparrow} ({\bf r}',\tau';{\bf r},\tau) +
G_{\bar\alpha\downarrow}({\bf r},\tau;{\bf r}',\tau')
F_{\alpha\uparrow} ({\bf r}',\tau';{\bf r},\tau) ] \Bigr \rbrace ,
\end{eqnarray}
where ${\bf q}=(q_x,q_y=0)$ and $\tilde q=(q_x,\omega_\nu)$. Using 
\begin{eqnarray}
\int d^2r\,d\tau\, \int d^2r'\,d\tau'\, e^{-i({\bf q}\cdot {\bf
r}-\omega_\nu \tau) + i({\bf q}'\cdot {\bf r}'-\omega_\nu' \tau') 
+ i\alpha {\bf Q}_{pN} \cdot {\bf r}'} 
G_{\alpha\uparrow}({\bf r},\tau;{\bf r}',\tau') 
F_{\bar\alpha\downarrow} ({\bf r}',\tau';{\bf r},\tau) = && \nonumber \\ 
\delta_{\tilde q,\tilde q'} I_{pN} \sum_{{\bf k},\omega} 
G_{\alpha\uparrow}({\bf k},\omega) 
F_{\bar\alpha\downarrow} ({\bf k}-\alpha {\bf Q}_0-{\bf
q},\omega-\omega_\nu) e^{i\alpha pN(k_yb-\pi/2)} , &&
\\
\int d^2r\,d\tau\, \int d^2r'\,d\tau'\, e^{-i({\bf q}\cdot {\bf
r}-\omega_\nu \tau) + i({\bf q}'\cdot {\bf r}'-\omega_\nu' \tau') 
+ i\alpha {\bf Q}_{pN} \cdot {\bf r}'} 
F_{\bar\alpha\downarrow}({\bf r},\tau;{\bf r}',\tau') 
G_{\bar\alpha\downarrow} ({\bf r}',\tau';{\bf r},\tau) = && \nonumber \\  
\delta_{\tilde q,\tilde q'} I_{pN} \sum_{{\bf k},\omega} 
F_{\bar\alpha\downarrow}({\bf k},\omega)
G_{\bar\alpha\downarrow} ({\bf k}-{\bf q},\omega-\omega_\nu)
e^{i\alpha pN(k_yb-\pi/2)} , && 
\\
\int d^2r\,d\tau\, \int d^2r'\,d\tau'\, e^{-i({\bf q}\cdot {\bf
r}-\omega_\nu \tau) + i({\bf q}'\cdot {\bf r}'-\omega_\nu' \tau') 
- i\alpha {\bf Q}_{pN} \cdot {\bf r}'} 
F_{\alpha\uparrow}({\bf r},\tau;{\bf r}',\tau') 
G_{\alpha\uparrow} ({\bf r}',\tau';{\bf r},\tau) = && \nonumber \\  
\delta_{\tilde q,\tilde q'} I_{pN} \sum_{{\bf k},\omega} 
F_{\alpha\uparrow}({\bf k},\omega)
G_{\alpha\uparrow} ({\bf k}-{\bf q},\omega-\omega_\nu)
e^{-i\alpha pN(k_yb-\pi/2)} , &&
\\
\int d^2r\,d\tau\, \int d^2r'\,d\tau'\, e^{-i({\bf q}\cdot {\bf
r}-\omega_\nu \tau) + i({\bf q}'\cdot {\bf r}'-\omega_\nu' \tau') 
- i\alpha {\bf Q}_{pN} \cdot {\bf r}'} 
G_{\bar\alpha\downarrow}({\bf r},\tau;{\bf r}',\tau') 
F_{\alpha\uparrow} ({\bf r}',\tau';{\bf r},\tau) = && \nonumber \\  
\delta_{\tilde q,\tilde q'} I_{pN} \sum_{{\bf k},\omega} 
G_{\bar\alpha\downarrow}({\bf k},\omega)
F_{\alpha\uparrow}({\bf k}+\alpha {\bf Q}_0-{\bf q},\omega-\omega_\nu)
e^{-i\alpha pN(k_yb-\pi/2)} , &&
\end{eqnarray}
we obtain
\begin{eqnarray}
\rho_{\rm DW}(\tilde q) &=& -i \sum_{\alpha,p} \tilde\Delta_{pN}
\tilde\varphi^{(pN)}_{\alpha\uparrow} (\tilde q) \frac{T}{L_xL_y} \sum_{{\bf
k},\omega} \nonumber \\ && \times \Bigl \lbrace 
e^{i\alpha pN(k_yb-\pi/2)} [ G_{\alpha\uparrow}({\bf k},\omega) 
F_{\bar\alpha\downarrow} ({\bf k}-\alpha {\bf Q}_0-{\bf
q},\omega-\omega_\nu) + F_{\bar\alpha\downarrow}({\bf k},\omega)
G_{\bar\alpha\downarrow} ({\bf k}-{\bf q},\omega-\omega_\nu) ] \nonumber \\ &&
- e^{-i\alpha pN(k_yb-\pi/2)} [ F_{\alpha\uparrow}({\bf k},\omega)
G_{\alpha\uparrow} ({\bf k}-{\bf q},\omega-\omega_\nu) +
G_{\bar\alpha\downarrow}({\bf k},\omega)
F_{\alpha\uparrow}({\bf k}+\alpha {\bf Q}_0-{\bf q},\omega-\omega_\nu) ]
\Bigr \rbrace .
\end{eqnarray}
To lowest order in $v_Fq_x$ and $\omega_\nu$, we have (for $q_y=0$)  
\begin{eqnarray}
\frac{T}{bL_x} \sum_{k_x,\omega} 
G_{\alpha\sigma}({\bf k},\omega) 
F_{\bar\alpha\bar\sigma}{\bf k}-\alpha {\bf Q}_0-{\bf q},\omega-\omega_\nu) 
&=& -N(0) \frac{\tilde\Delta_{\alpha\uparrow}^*(k_y)}{8|\tilde\Delta
_{\alpha\uparrow}(k_y)|^2} (i\omega_\nu +\alpha v_Fq_x) , \\ 
\frac{T}{bL_x} \sum_{k_x,\omega} 
F_{\bar\alpha\bar\sigma}({\bf k},\omega) 
G_{\bar\alpha\bar\sigma}{\bf k}-{\bf q},\omega-\omega_\nu) 
&=& -N(0) \frac{\tilde\Delta_{\alpha\uparrow}^*(k_y)}{8|\tilde\Delta
_{\alpha\uparrow}(k_y)|^2} (-i\omega_\nu +\alpha v_Fq_x) .
\end{eqnarray}
This leads to
\begin{equation}
\rho_{\rm DW} (\tilde q) = \frac{iN(0)}{4N_\perp} \sum_{p,\alpha,k_y} \tilde
\Delta_{pN} \alpha v_F q_x e^{i\alpha pN(k_yb-\pi/2)} 
\frac{\tilde\Delta_{\alpha\uparrow}^*(k_y)}{|\tilde\Delta
_{\alpha\uparrow}(k_y)|^2} \Bigl ( \tilde\varphi^{(pN)}_{\alpha\uparrow}
(\tilde q) - \tilde\varphi^{(pN)}_{\bar\alpha\uparrow} (\tilde q) \Bigr ) .
\end{equation}
Performing the sum over $k_y$, we finally obtain
\begin{equation}
\rho_{\rm DW} (\tilde q) = i \frac{q_x}{2\pi b}  \Bigl (
\tilde\varphi^{(pN)}_{+\uparrow} (\tilde q) -
\tilde\varphi^{(pN)}_{-\uparrow} (\tilde q) \Bigr ) ,
\end{equation}
which yields Eq.~(\ref{charge1}).

\section{Mean-field susceptibilities in the helicoidal phase}
\label{app:hel}

In the helicoidal phase, the  mean-field susceptibilities are given by
\begin{eqnarray}
\bar\chi_{\alpha\sigma,\alpha\sigma} ({\bf q}+\alpha {\bf
Q}_{p(\alpha,\sigma)N}, {\bf q}+\alpha {\bf
Q}_{p(\alpha,\sigma)N}, \omega_\nu) &=& 
I_{p(\alpha,\sigma)N}^2
\frac{N(0)}{2} \biggl [ \ln \frac{2E_0}{|\tilde\Delta_{\alpha\sigma}|}
-\frac{1}{2} - \frac{\omega^2_\nu + v_F^2q_x^2}{6
\tilde\Delta^2_{\alpha\sigma}} \biggr ] ,  
\label{chi10} \\ 
\bar\chi_{\alpha\sigma,\bar\alpha\bar\sigma} ({\bf q}+\alpha {\bf
Q}_{p(\alpha,\sigma)N}, {\bf q}-\alpha {\bf Q}_{p(\alpha,\sigma)N}, 
\omega_\nu) &=&  
I_{p(\alpha,\sigma)N}^2 \frac{N(0)}{2} \biggl [
-\frac{1}{2} + \frac{\omega^2_\nu + v_F^2q_x^2}{12
\tilde\Delta^2_{\alpha\sigma}} \biggr ] ,
\end{eqnarray}
for $q_y=0$ and $|\omega_\nu|,v_F|q_x|\ll |\tilde\Delta_{\alpha\sigma}|$. 
Introducing the notations 
\begin{eqnarray}
c_{++} &=& g_2 [ \bar\chi_{+\uparrow,+\uparrow} ({\bf q}+{\bf Q}_N, {\bf
q}+{\bf Q}_N,\omega_\nu) 
- \bar\chi_{+\uparrow,-\downarrow} ({\bf q}+{\bf Q}_N,{\bf q}-{\bf Q}_N,
\omega_\nu) ] , \nonumber \\ 
c_{--} &=& g_2 [ \bar\chi_{-\uparrow,-\uparrow} ({\bf q}-{\bf Q}_{\bar N},
{\bf q}-{\bf Q}_{\bar N}, \omega_\nu) 
- \bar\chi_{-\uparrow,+\downarrow} ({\bf q}-{\bf Q}_{\bar N},{\bf
q}+{\bf Q}_{\bar N},\omega_\nu) ] ,
\end{eqnarray}
the gap equation reads
\begin{equation}
(1-\bar c_{++}) (1-\bar c_{--}) -r^2 \bar c_{++} \bar c_{--} =0 ,
\end{equation}
where
$\bar c_{++} = 1/(1+\gamma r)$ and $\bar c_{--} = \gamma/(\gamma+r)$  
($\bar c_{pp}=c_{pp}|_{\tilde q=0}$).

\eleq

\bleq

\begin{figure}
\epsfysize 14.cm 
\epsffile[50 200 395 750]{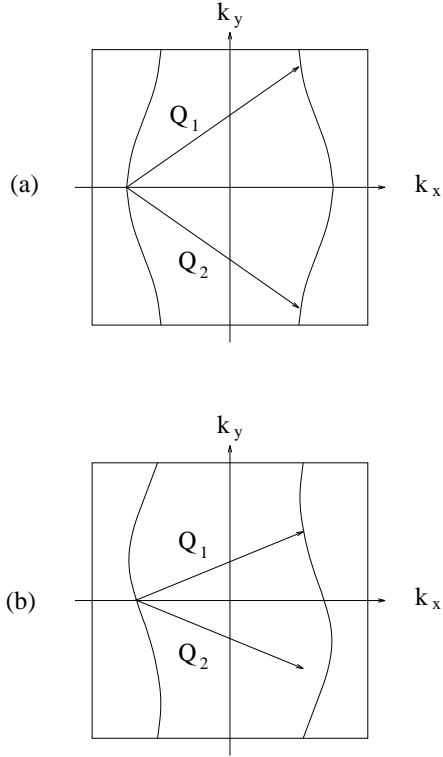}
\caption{(a) Fermi surface deduced from the dispersion law (\ref{E0}) with
$\kappa =0$. At half-filling, the two `best' nesting vectors ${\bf Q}_1$ and
${\bf Q}_2$ are coupled by umklapp scattering. (b) Fermi surface for $\kappa
=\pi /4$. There is only one `best' nesting vector (${\bf Q}_1$). By umklapp
scattering, it couples to ${\bf Q}_2$, which is not a good nesting vector.}
\label{fig:FS}
\end{figure}

\begin{figure}
\epsfysize 13.cm 
\epsffile[0 400 395 850]{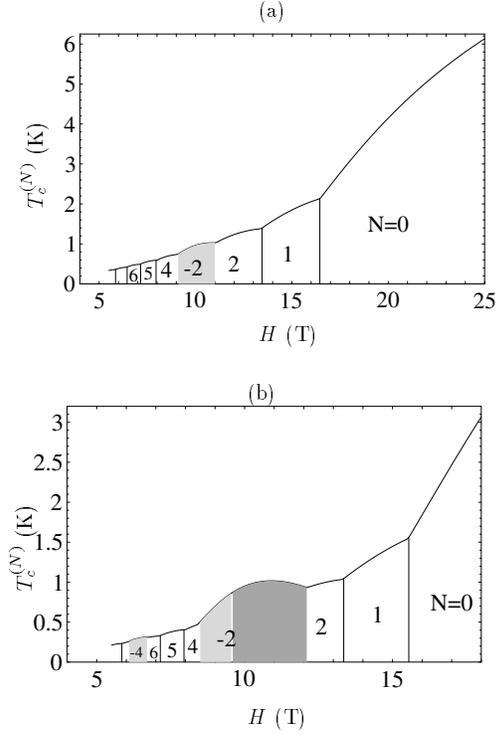}
\caption{Transition temperature $T_c^{(N)}$ between the metallic phase and
the FISDW phases in presence of umklapp processes. The vertical
lines are only guides for the eyes and do not necessarily correspond to
the actual first-order transition lines between FISDW phases. 
(a) $g_3/g_2=0.03$. [$g_3$
and $g_2$ are the strengths of normal and umklapp processes, respectively
(see Sec.~\ref{sec:model}).] The shaded area corresponds to the Ribault
phase $N=-2$. (b) $g_3/g_2=0.06$. Two negative phases, $N=-2$ and $N=-4$, are
observed (shaded areas). The phase $N=-2$ splits into two subphases:
helicoidal (dark shaded area) and sinusoidal (light shaded area). }
\label{fig:Tc}
\end{figure}

\begin{figure}
\epsfysize 6.cm 
\epsffile[0 500 330 705]{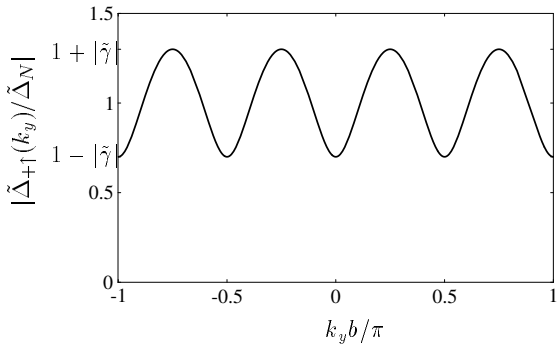}
\caption{Quasi-particle excitation gap $2|\tilde\Delta_{+\uparrow}(k_y)|$
in the phase $N=-2$ for $|\tilde\gamma|=1/3$. }
\label{fig:QP}
\end{figure}

\begin{figure}
\epsfysize 6.cm 
\epsffile[0 505 330 725]{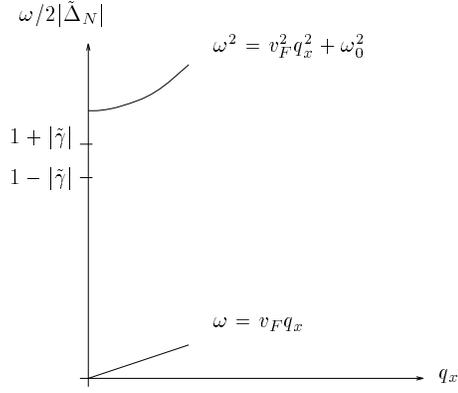}
\caption{Dispersion laws of the two sliding modes in the Ribault phase. The
gapped mode is 
generally expected to lie above the quasi-particle excitation gap. The
latter varies between $2|\tilde\Delta_N|(1-|\tilde\gamma|)$ and
$2|\tilde\Delta_N|(1+|\tilde\gamma|)$ (Fig.~\ref{fig:QP}).}
\label{fig:cm}
\end{figure}

\begin{figure}
\epsfysize 5.5cm 
\epsffile[0 490 330 685]{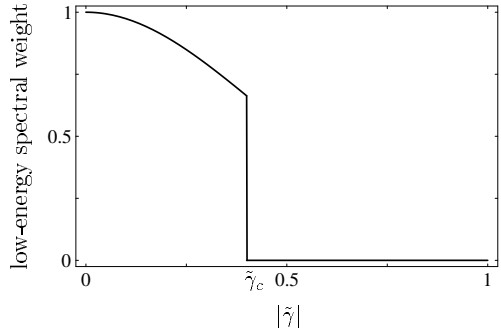}
\caption{Fraction of the total optical spectral weight $\omega_p^2/4$
carried by the low-energy collective modes. In the sinusoidal phase
($|\tilde\gamma| \leq \tilde\gamma_c \sim 0.4$), the low-energy sliding mode
carries the fraction  $3(1-\tilde\gamma^2)/(3+5\tilde\gamma^2)$ of the total
spectral weight. In the helicoidal phase, there is no spectral weight at low
energy (see Sec.~\ref{sec:hel}). } 
\label{fig:sw}
\end{figure} 

\begin{figure}
\epsfysize 5.5cm 
\epsffile[0 490 330 685]{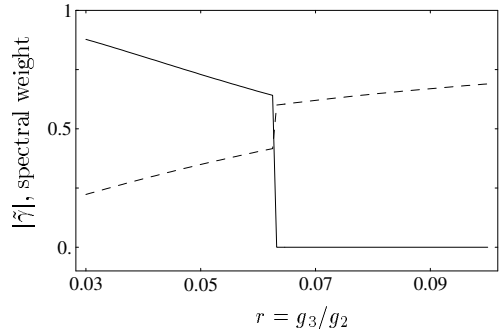}
\caption{$|\tilde\gamma|$ (dashed line) and low-energy optical spectral
weight (solid line) as a function of $r=g_3/g_2$ in the phase $N=-2$ ($H=10$
T). The transition from the
sinusoidal to the helicoidal phase occurs for $r\sim 0.06$. }
\label{fig:sw1}
\end{figure}

\begin{figure}
\epsfysize 6.cm 
\epsffile[100 375 430 600]{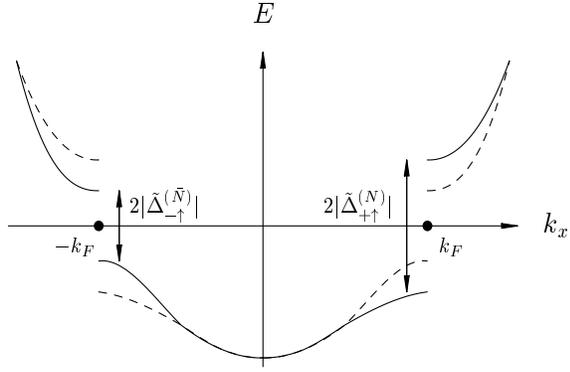}
\caption{Quasi-particle excitation spectrum in the helicoidal phase. The
solid (dashed) line corresponds to up (down)  spins. For clarity, we have
not shown the Zeeman splitting. }
\label{fig:spec}
\end{figure}

\eleq 

\ecols


\begin{references}
\bibitem{note8} The condition ${\bf Q}_1\neq {\bf Q}_2$ does not necessary
imply that the two SDW's are incommensurate. Nevertheless, we assume here
for simplicity that this is the case. This assumption is correct for a
system closed to half-filling.
\bibitem{note9} The longitudinal components of the nesting vectors may also
be different from $2k_F$. This is however not important for our qualitative
discussion.
\bibitem{note0} Note that the Bechgaard salts have a triclinic
crystallographic structure. Nevertheless, it is believed that the
orthorhombic Fermi surface shown in Fig.~\ref{fig:FS}b is a good
approximation to their actual Fermi surface as far as the nesting properties
are concerned.
\bibitem{note10} The precise criterion is $t_{2b}t_{4b}>0$, where $t_{2b}$
and $t_{4b}$ are higher harmonics in the transverse dispersion law (see
Sec.~\ref{sec:model}).
\bibitem{note3} Note that the definition of the order parameters $\Delta
^{(pN)}_{\alpha\sigma}$ slightly differs from Ref.~\onlinecite{Dupuis98}.
\bibitem{note6} The mean-field equations indicate that $\gamma$ varies
between $-r$ and $1$ and can therefore be negative. In the numerical
calculation, we find that $\gamma\gg r$ is positive in the Ribault phase.
\bibitem{note4} Note that for a helicoidal SDW, one cannot distinguish
between a uniform spin rotation and a global translation.
\end{references}
\end{document}